\documentclass[fleqn,10pt]{wlscirep}
\usepackage[utf8]{inputenc}
\usepackage[T1]{fontenc}
\usepackage{array}
\usepackage{geometry}
\usepackage{soul}
\usepackage{adjustbox}

\title{Easy-access online social media metrics are associated with misinformation sharing activity}

\author[1]{Júlia Számely}
\author[2]{Alessandro Galeazzi}
\author[3,4]{Júlia Koltai}
\author[1,*]{Elisa Omodei}
\affil[1]{Department of Network and Data Science, Central European University, Vienna, Austria}
\affil[2]{Department of Mathematics, University of Padova, Padova, Italy}
\affil[3]{MTA–TK Lendület “Momentum” Digital Social Science Research Group for Social Stratification, HUN-REN Centre for Social Sciences, Budapest, Hungary}
\affil[4]{Department of Social Research Methodology, Faculty of Social Sciences, Eötvös Loránd University, Budapest, Hungary}

\affil[*]{omodeie@ceu.edu}


\begin{abstract}
Misinformation poses a significant challenge studied extensively by researchers, yet acquiring data to identify primary sharers is time-consuming and challenging. To address this, we propose a low-barrier approach to differentiate social media users who are more likely to share misinformation from those who are less likely. Leveraging insights from previous studies, we demonstrate that easy-access online social network metrics—average daily tweet count, and account age—can be leveraged to help identify potential low factuality content spreaders on X (previously known as Twitter). We find that higher tweet frequency is positively associated with low factuality in shared content, while account age is negatively associated with it. We also find that some of the effects differ depending on the number of accounts a user follows. Our findings show that relying on these easy-access social network metrics could serve as a low-barrier approach for initial identification of users who are more likely to spread misinformation, and therefore contribute to combating misinformation effectively on social media platforms.
\end{abstract}

\begin{document}

\flushbottom
\maketitle
%
%
\thispagestyle{empty}


\section*{Introduction}
The problem of the global spread of misinformation has been recognised as one of the most pressing issues in today’s societies, prompting many researchers to dedicate efforts to understanding the issue as well as to developing potential solutions for tackling it \cite{WEF2024}. Researchers have focused on a variety of areas to understand misinformation, examining the phenomenon from the perspectives of the misinformation producers, consumers, the messages themselves, and the contexts in which the messages are embedded \cite{Chen2023}. While bots may play a significant role in spreading misinformation, the role of human users in sharing it remains a key element in the diffusion process \cite{Vosoughi2018}. Furthermore, as the spreading of misinformation carries the risk of subjecting people to manipulation and might influence behaviour in undesirable ways \cite{Ferrara2020, Linden2024}, understanding the factors that affect individuals’ likelihood of sharing misinformation is essential.

The reasons behind people’s tendency to share misinformation have been widely studied, and findings have shown that in many cases the decision of users to share misinformation does not stem from malicious intent, but rather reflects a failure to consider the accuracy of the piece of news shared \cite{Pennycook2021}. Two pathways have been proposed for this failure: limited attention to accuracy in the moment of sharing, or difficulty in distinguishing between true and false content \cite{Pennycook2021}. In particular, even individuals who can distinguish between true and false information more easily often fail to apply this discernment at the point of sharing, due to contextual or cognitive factors. First, when sharing content on social media, individuals are prone to prioritising social considerations, such as how sharing certain content affects their social status. Users might also use the content they share as a way to connect to others \cite{Apuke2021, Islam2020, Balakrishnan2021}. Second, limited attention caused by high information volume on social media platforms reduces users’ ability to reflect on the accuracy of each piece of content they encounter \cite{Apuke2022}. These processes have been linked to various individual-level traits, including limited cognitive skills \cite{Arechar2023}, lack of analytical reasoning \cite{Pennycook2021b}, and lower levels of digital literacy \cite{Guess2020, fainmesser2021}.

Interventions on social media have been shown to be successful across various contexts. Examples include prompting users to think about accuracy \cite{Arechar2023}, and ‘inoculating’ users with small doses of misinformation and then refuting that piece of misinformation to increase awareness of how one can be manipulated into trusting content \cite{Rozenbeek2022}. Given their success, it is compelling to explore ways in which social media users who are more likely to engage with misinformation sharing could be targeted with such interventions. While previous research presented above provides ample evidence on where to look for identifying individuals prone to misinformation sharing, obtaining data on these predictor characteristics can be complicated, time-consuming, and/or expensive \cite{Tucker2018}. 

In this paper, we set out to provide a low-barrier, explanatory approach to assess whether simple, easily accessible account-level characteristics are systematically associated with users’ likelihood of sharing misinformation, drawing on insights from previous literature and using easy-access data from X (Twitter). Our aim in utilising X metrics as accessible proxies for the above outlined phenomena is to explore whether they can serve as early-warning signals of lower factuality tendencies among users. We aim to show that basic information available on X accounts can help surface broad patterns associated with lower factuality, offering a potential foundation for scalable and low-cost interventions. 

For this purpose, we selected four metrics to test: (1) follower count, (2) average daily tweet count, (3) followed account count, and (4) account age. In our criteria for selecting these metrics we prioritised metrics that are widely available across platforms; not reliant on user self-report, thereby reducing biases associated with self-disclosure; and retrievable in a structured and standardized way, facilitating potential cross-platform generalizability; as well as the metrics' potential ability to serve as proxies for the significant determinants of misinformation sharing found in the literature.

While we focus on these metrics due to their accessibility, we recognize that other account-level characteristics may also be useful for understanding users’ likelihood of sharing low-factuality content. For example, features such as whether a user provides a location or the complexity of their bio text could add additional explanatory power. Additionally, we acknowledge that alternative methods—such as timeline content or network-based analysis—may yield greater accuracy in detecting misinformation \cite{Truong2024}. However, these approaches typically require substantially more data collection and processing effort, or more advanced computation. In contrast, the metrics we rely on—follower and following counts, tweet frequency, and account age—are embedded in the basic profile metadata and can be obtained and analyzed with minimal technical and infrastructural requirements. This makes them especially valuable in contexts where data access is restricted, resources are limited, or scalability is a priority. In other words, our metric selection was informed by the literature to ensure efficiency, however we prioritised data accessibility instead of the selection of perfect proxies. 

Based on these criteria, we chose the following metrics to represent three lines of misinformation sharing determinants: follower count for social considerations, average daily tweet count and number of users followed for cognitive overload, and account age for digital literacy. 
Our hypothesis for each metric is that it could serve as a useful indicator for distinguishing between low and high factuality users, with specific expectations for how each metric would affect user factuality. We expect to find users with more followers and the ones who tweet frequently and follow more accounts to be less factual, while users with older accounts are expected to be more factual. The expectation regarding follower count is based on research suggesting that social motivations—such as maintaining reputation or appealing to an audience—can divert attention from content accuracy \cite{Pennycook2021}. A higher number of followers may increase perceived audience pressure or reputational concerns, which have been shown to influence sharing decisions. 
Similarly, we hypothesize that users who produce large volumes of tweets or follow many accounts experience a form of cognitive strain or information overload, reducing the attention they can dedicate to assessing accuracy \cite{Apuke2022, Qiu2017}.
We acknowledge that platform algorithms shape users’ content exposure, and that a user's feed may include messages from both followed and non-followed accounts. Nonetheless, following more accounts likely increases the potential content diversity and volume a user may encounter, even within algorithmically filtered feeds.
Finally, we interpret account longevity as a rough proxy for digital experience or literacy. More precisely, we treat account age as a behavioral approximation of platform familiarity and accumulated exposure, rather than digital literacy in a strict sense. We also note that account age may correlate with user age or other demographic traits, which complicates its interpretation. Account longevity could reflect a mix of factors — including but not limited to digital literacy — such as habitual platform use, early adopter behavior (which has been linked to higher media savvy \cite{Kelleher2012}), and greater exposure to platform norms and misinformation corrections. This interpretation aligns with prior work on digital behavior and literacy \cite{Hargittai2005}, while also acknowledging the conceptual limitations of using account age as a stand-in for digital skills.
Hence, although several techniques exist to infer users' propensity to consume unreliable sources, the development of a method based on easily accessible metrics—grounded in the literature on misinformation susceptibility—may offer a useful framework for understanding broad patterns of vulnerability and the design of targeted countermeasures to curb the spread of false or inaccurate information.

\section*{Results}
We analysed data collected from X in 2022 (at that time still known as Twitter), consisting of 1,670,127 users and 14,688,374 tweets. Each user in the dataset is assigned a factuality score based on the credibility of the news sources they share, determined from their last 500 tweets containing a URL. Using these scores, we categorised users into high, middle, and low factuality groups (see Methods). Additionally, we retrieved various social network metrics on the same users, including follower count, number of followed accounts, tweet count, and registration date. Finally, we created 1,000 randomised versions of the dataset, where we shuffled users’ factuality scores while keeping all other features fixed, to serve as a basis for testing the statistical significance of our results (see Methods for more details on the dataset).

In the following, we present our findings on easy-access social network metrics’ effectiveness in the initial identification of individuals who tend to share a large amount of misinformation. We draw on the literature introduced above to guide the selection of the social network metrics tested. Specifically, drawing upon three strands of relevant literature – on social motivations, limited attention, and digital literacy – we select four metrics: follower count, daily average tweet count, followed account count, and the number of days since registration. 

To ensure that our analysis focuses on regular users rather than automated or public accounts (e.g., news sites, celebrities), we applied specific filtering thresholds. In our main analysis, we used the following middle-level filtering criteria: only non-verified accounts were included; accounts were limited to a maximum of 32 tweets per day on average; follower count was capped at 10,000; follower-to-following ratio had to be 5 or lower; followed account count was capped at 10,000; following-to-follower ratio had to be 10 or lower. We report these filtering criteria and their justification in \textit{Filtering the dataset for regular users} and Table \ref{table:table1}.

To construct low and high factuality groups, we divided users into factuality tertiles based on their domain-sharing profiles—focusing on the bottom and top 30\% in the main analysis, and testing 25\% and 35\% thresholds for robustness checks (see Supplementary Information).

\subsection*{Popularity is associated with lower factuality}
\begin{figure}
      \centering
      \includegraphics[width=.7\textwidth]{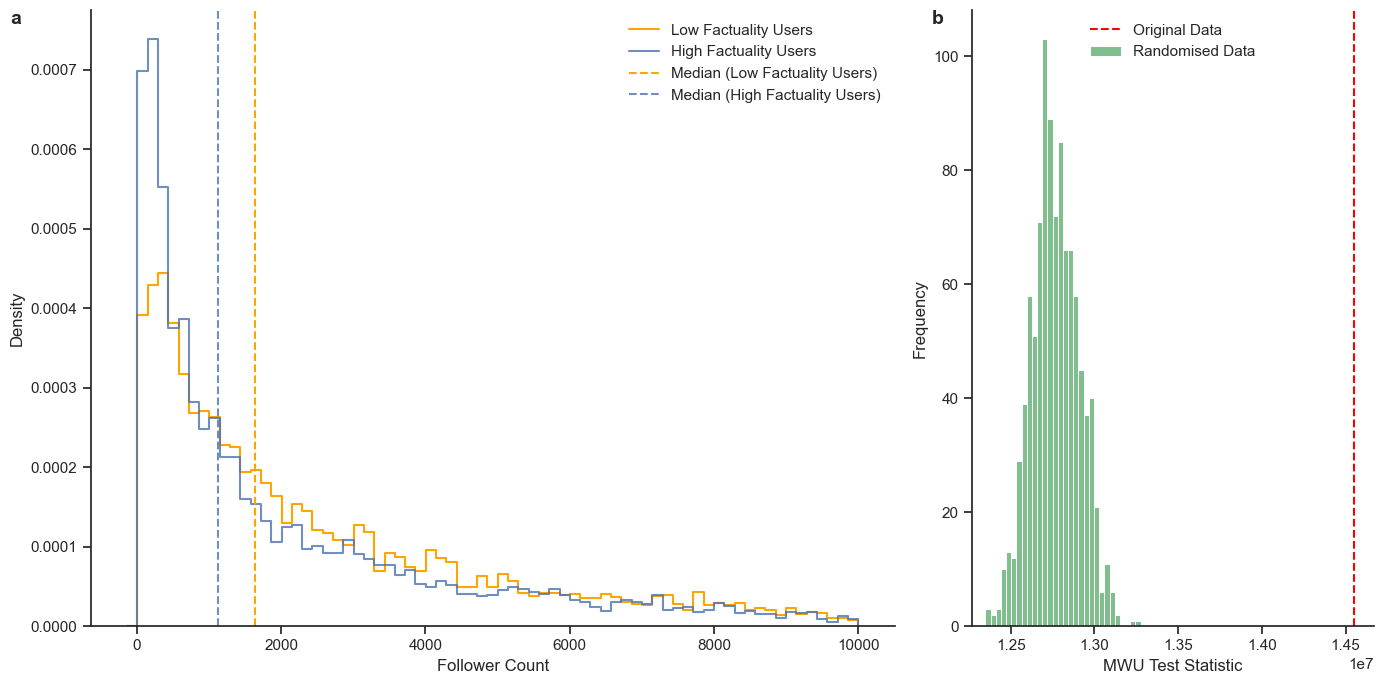}
      \caption{\textbf{Low factuality users tend to have higher follower count than high factuality users.} Panel a: Comparing the distributions of follower count between low (orange curve) and high factuality (blue curve) users, we find that low factuality users have a significantly higher number of followers compared to high factuality users (p-value $<$ 0.0001). The median value for low factuality users (orange dotted line) is 1642 followers, whereas for high factuality users (blue dotted line) it is 1125.   Panel b: The MWU test score obtained from the empirical data (red dotted line) is higher than all the MWU test scores calculated on the 1,000 shuffled datasets (green bars).}
      \label{fig:fig1}
    \end{figure}

Building on the literature discussed above, highlighting the dominance of social considerations in social media sharing over accuracy concerns \cite{Pennycook2021}, we selected follower count on X (popularity) as our first metric. We hypothesise that follower count on X will serve as an indicator to identify users more inclined to share misinformation.
To this end, we employed one-sided Mann-Whitney U (MWU) tests to compare the follower count distributions of the low and high factuality users, and verified their statistical significance by comparing them with the MWU statistics obtained on the 1,000 randomised datasets.
Low factuality users were on average found to have a higher number of followers compared to high factuality ones, confirming our initial expectations (\hyperref[fig:fig1]{Figure 1}a). The results were found to be statistically significant by comparing the MWU test statistic with the MWU test statistics obtained using the 1000 shuffled datasets (\hyperref[fig:fig1]{Figure 1}b). This finding thus suggests that follower count can serve as an initial indicator in distinguishing between low and high factuality users, confirming our hypothesis.

\subsection*{High tweeting rate and followed account count are also associated with lower factuality}
\begin{figure}
      \centering
      \includegraphics[width=.7\textwidth]{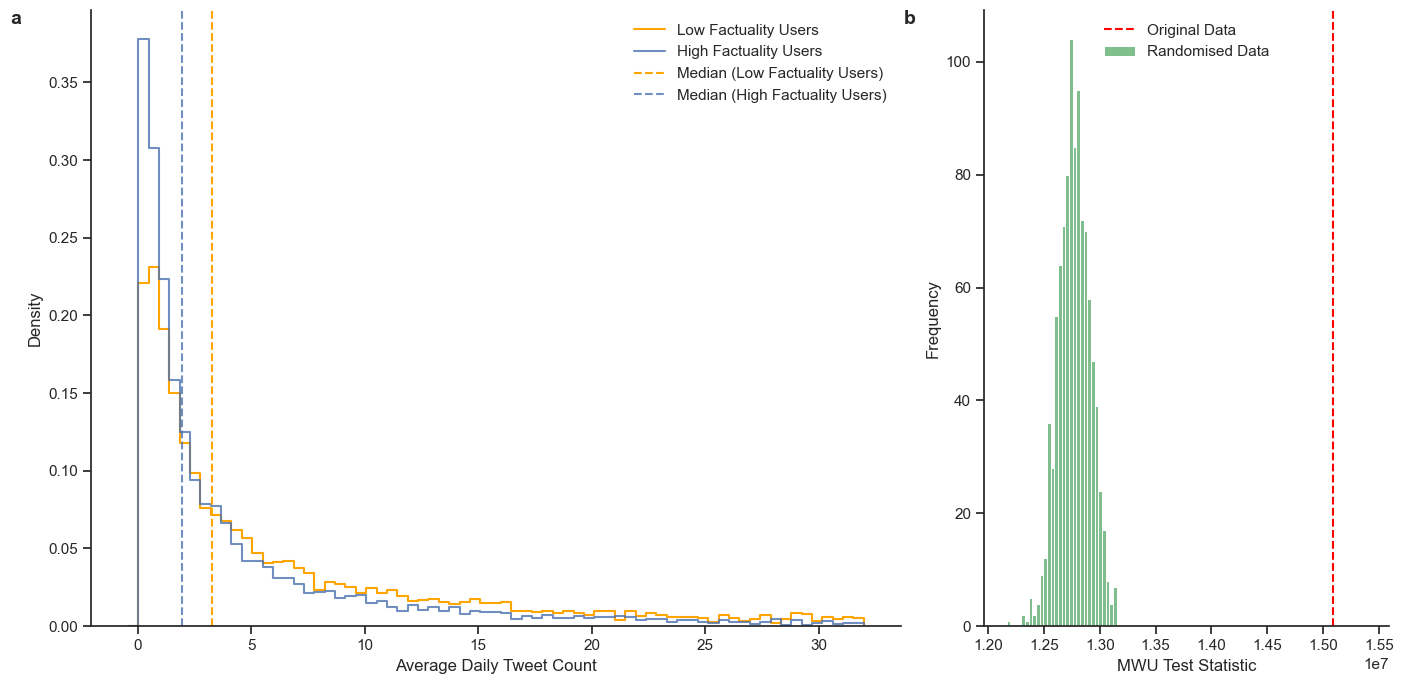}
      \caption{\textbf{Low factuality users tend to have higher tweet count than high factuality users.} Panel a: Comparing the distributions of tweet count between low (orange curve) and high factuality (blue curve) users, we find that low factuality users have a significantly higher number of tweets compared to high factuality users (p-value $<$ 0.0001). The median value for low factuality users (orange dotted line) is 3.26 tweets per day, whereas for high factuality users (blue dotted line) it is 1.94. Panel b: The MWU test score obtained from the empirical data (red dotted line) is higher than all the MWU test scores calculated on the 1,000 shuffled datasets (green bars).}
      \label{fig:fig2}
    \end{figure}

\begin{figure}
      \centering
      \includegraphics[width=.7\textwidth]{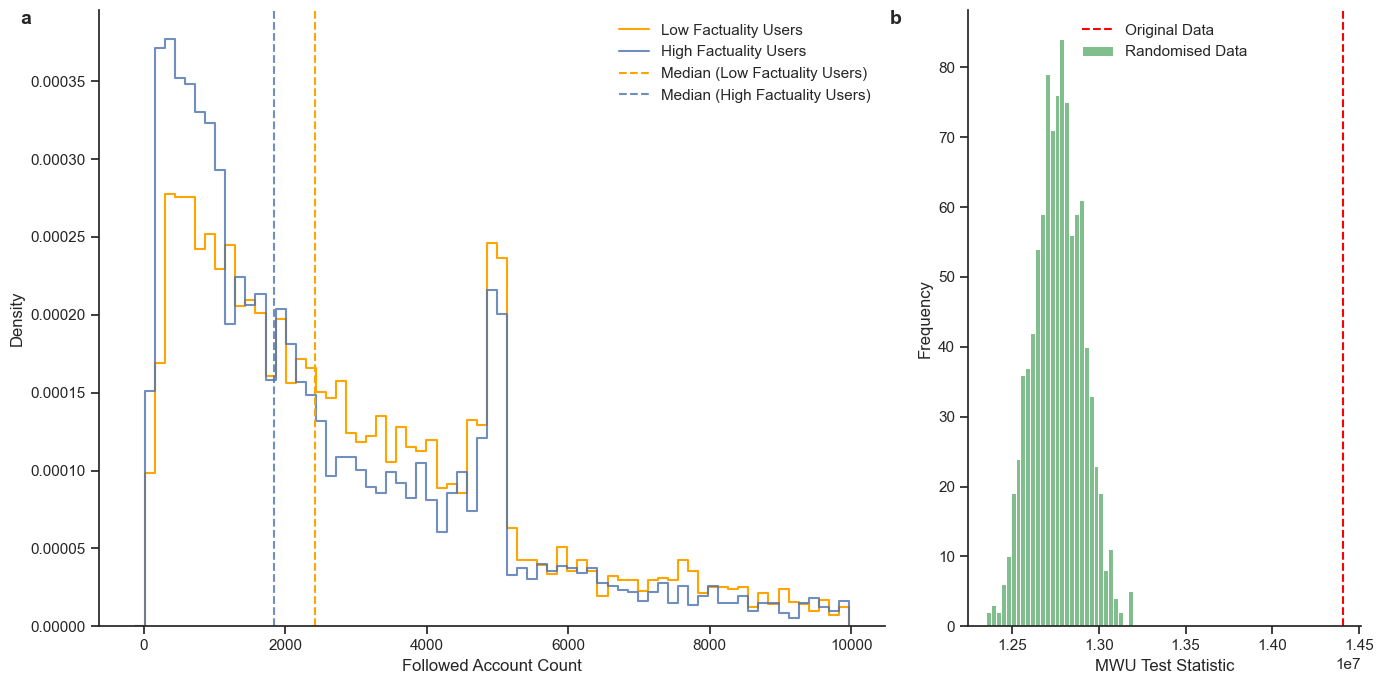}
      \caption{\textbf{Low factuality users tend to have higher followed account count than high factuality users.} Panel a: Comparing the distributions of followed account count between low (orange curve) and high factuality (blue curve) users, we find that low factuality users have a significantly higher number of followed accounts compared to high factuality users (p-value $<$ 0.0001). The median value for low factuality users (orange dotted line) is 2415 followed accounts, whereas for high factuality users (blue dotted line) it is 1845. The peak around 5000 is due to an X policy that limits the number of new followed accounts until the user obtains more followers. Panel b: The MWU test score obtained from the empirical data (red dotted line) is higher than all the MWU test scores calculated on the 1,000 shuffled datasets (green bars).}
      \label{fig:fig3}
    \end{figure}

Based on the related literature on information overload \cite{Apuke2022}, we selected the number of tweets per day and the number of accounts followed as two additional metrics for distinguishing between low and high factuality users, and hypothesise that these two metrics are useful in distinguishing between low and high factuality users.
Similarly to the previous metric, we utilised MWU scores to compare the distributions of these two metrics between low and high factuality users, and validated their significance against the 1,000 shuffled datasets. 
Low factuality users were observed both to post a higher number of tweets per day (\hyperref[fig:fig2]{Figure 2}a) and to follow a greater number of users compared to their high factuality counterparts (\hyperref[fig:fig3]{Figure 3}a). Both results were found to be statistically significant by comparing the MWU test statistics obtained from the empirical data to the MWU test statistics obtained from the 1000 shuffled datasets (see \hyperref[fig:fig2]{Figure 2}b and \hyperref[fig:fig2]{Figure 3}b, respectively). These results suggest that the two metrics tested can be useful in distinguishing between low and high factuality users.

\subsection*{Longer social media presence is associated with higher factuality}
\begin{figure}
      \centering
      \includegraphics[width=.7\textwidth]{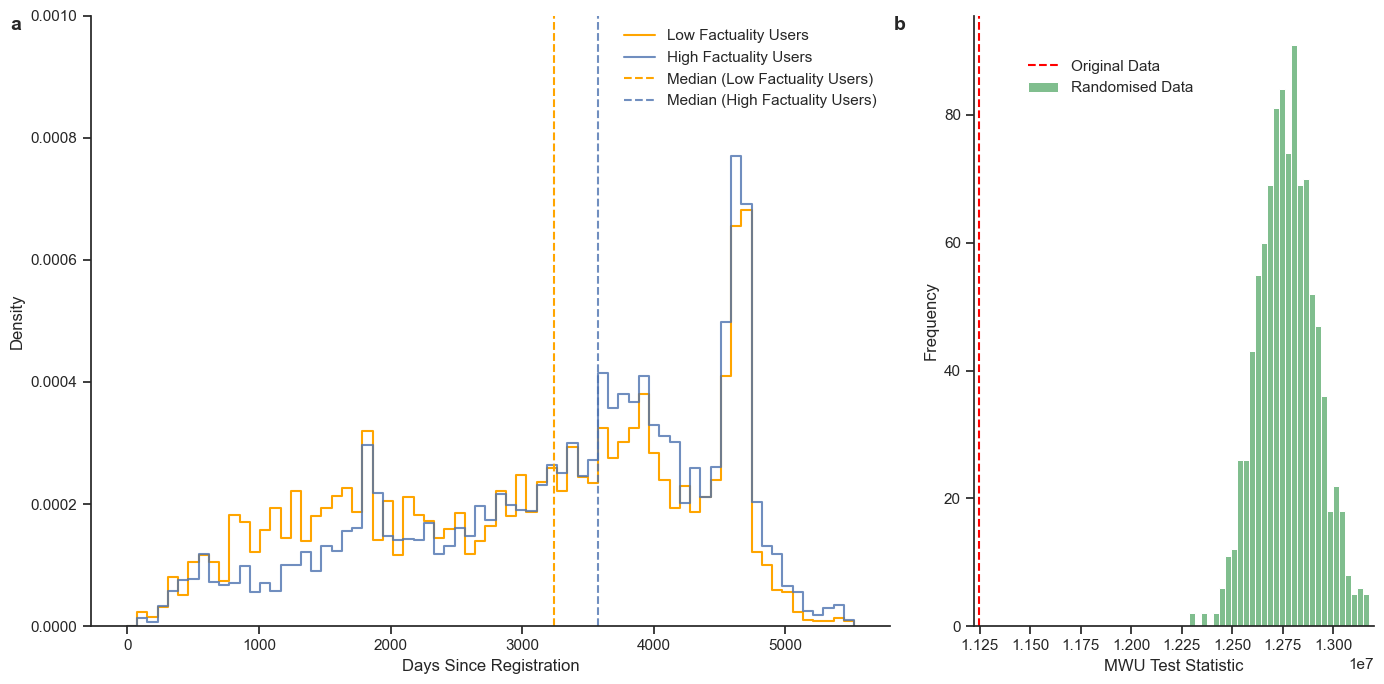}
      \caption{\textbf{Low factuality users tend to have lower number of days since registration than high factuality users.} Panel a: Comparing the distributions of the number of days since registration between low (orange curve) and high factuality (blue curve) users, we find that low factuality users have a significantly lower number of days compared to high factuality users (p-value $<$ 0.0001). The median value for low factuality users (orange dotted line) is 3242 days since registration, whereas for high factuality users (blue dotted line) it is 3578. Panel b: The MWU test score obtained from the empirical data (red dotted line) is lower than all the MWU test scores calculated on the 1,000 shuffled datasets (green bars).}
      \label{fig:fig4}
    \end{figure}

Based on the literature on the relationship between digital literacy and misinformation sharing behaviour \cite{Guess2020}, we selected the number of days since registration as our last key metric for distinguishing between low and high factuality users.
We employed MWU scores to compare distributions and ensured their statistical significance through comparisons with the 1,000 shuffled datasets.
Low factuality users displayed a lower number of days since registration compared to high factuality users, i.e. older accounts tend to be more factual, aligning with our expectation (\hyperref[fig:fig4]{Figure 4}a). The results were found to be statistically significant by comparing the MWU test statistics obtained from the empirical data to the MWU test statistics obtained from the 1000 shuffled datasets (\hyperref[fig:fig4]{Figure 4}b). This finding supports the relevance of this metric in distinguishing between low and high factuality users.

\subsection*{Combined effect of social network characteristics on factuality}
Next, we employed multinomial regression to move from analysing the associations of factuality with individual social network features separately to considering the features all together. We do this in two stages. First, we perform multinomial regression with no interactions between the independent variables, second, we add interactions between explanatory variables. With conducting regression analysis, we are able to examine the effect of each key metric while simultaneously controlling for the effects of all other metrics. Thus, these models help us analyse the interplay of the metrics on factuality, and evaluate the extent to which these social network characteristics are associated with factuality scores.
    
We considered factuality as the outcome variable with three levels—low, middle, and high factuality. To capture the factors associated with notably low or high factuality, in each analysis we chose middle factuality to be the reference group that the other two are compared to. Explanatory variables included the four X metrics used above: follower count, followed account count, tweets per day, and days since registration. The explanatory variables were standardised for comparability.
To measure the effects, we assessed the average marginal effects (AMEs) of the four social network metrics, as well as interactions between them. According to the results, each of the four social network metrics was found to have a significant relationship with both the low and high factuality groups—compared to the middle factuality group—with a number of significant interactions between variables.
    
\begin{figure}
      \centering
      \includegraphics[width=.7\textwidth]{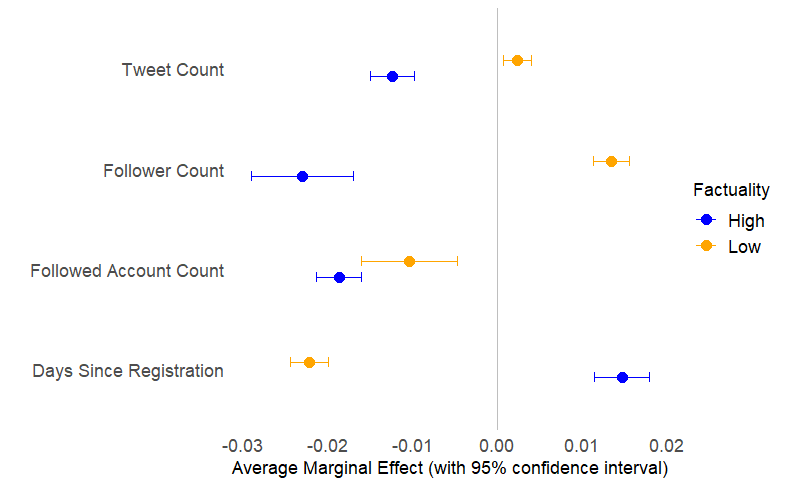}
      \caption{\textbf{Average marginal effects of social network characteristics on factuality.} All examined social network metrics are significantly associated with factuality, and the differences in the effects between low factuality (marked with orange) and high factuality (marked with blue) users is significant in all four cases. Zero effect would mean users are equally likely to be low or high factuality as middle factuality. Tweet count: higher tweet count means the user is more likely to be low factuality, and less likely to be high factuality than middle factual. Followed account count: higher followed account count means the user is less likely to be low factuality, as well as to be high factuality, but the latter effect is stronger. Follower count: higher follower count means the user is more likely to be low factuality, and less likely to be high factuality. Days Since Registration: higher number of days means the user is more likely to be high factuality, and less likely to be low factuality.}
      \label{fig:fig5}
    \end{figure}

First, we consider and report the AMEs of the four social network metrics on factuality (\hyperref[fig:fig5]{Figure 5}). Examining the AMEs of follower count on factuality group membership, we find that the more followers users have, the more likely they are to be in the low factuality group and less likely to be in the high factuality group, compared to the middle factuality group.
Specifically, a one standard deviation increase in follower count is associated with a 1.3 percentage point increase on average in the probability of belonging to the low factuality group, and on average a 2.3 percentage point decrease in the probability of belonging to the high factuality group, compared to the middle factuality group.
This suggests that, even when accounting for other variables, follower count remains a useful indicator for distinguishing between high and low factuality users.

The number of users followed is negatively associated with both the high and low factuality groups, meaning that holding all other variables constant, following a high number of accounts makes a user more likely to be in the middle factuality category, while it makes them less likely to be in the low or high factuality categories.
Quantitatively, a one standard deviation increase in the number of followed accounts is associated with a 1.0 percentage point decrease on average in the probability of being in the low factuality group and on average a 1.9 percentage point decrease in the probability of being in the high factuality group, compared to the middle factuality group.
These findings are on the one hand consistent with our expectations, as we see a negative association between the number of users followed and high factuality, compared to middle factuality. Opposite to our expectation, we also see a negative association between the number of users followed and belonging to the low factuality category compared to belonging to the middle factuality category, however this effect is significantly weaker than the one belonging to high factuality. These effects show that there is a reverse U-shaped relationship between followed account count and the three factuality groups, both with low and high factuality users typically following fewer users than middle factuality users. These results suggest that the power of followed account count as a measure to classify low and high factuality users is therefore limited. 
    
The average number of tweets produced by the user per day is negatively associated with being in the high factuality group compared to the middle factuality group, i.e. holding all other variables constant, high factuality users tweet less on average. For the low factuality group, we find a positive association compared to the middle factuality group, i.e. holding all other variables constant, low factuality users tweet more on average.
More specifically, a one standard deviation increase in average daily tweet count is associated with a 0.23 percentage point increase on average in the probability of being in the low factuality group, and on average a 1.2 percentage point decrease in the probability of being in the high factuality group, compared to the middle factuality group.
These results are in line with our hypothesis and previous results, confirming that average daily tweet count is a useful indicator in distinguishing between low and high factuality users. 
    
For days since registration, we see a positive effect for the high factuality, and the opposite for the low factuality group compared to the middle one.
Numerically, a one standard deviation increase in days since registration is associated with a 2.2 percentage point decrease on average in the probability of being in the low factuality group, and on average a 1.5 percentage point increase in the probability of being in the high factuality group, compared to the middle factuality group.
The results are therefore consistent with the outlined expectations and hypothesis, indicating that, holding all other variables constant, longer presence on X is associated with higher factuality, and that the metric is useful in distinguishing between low and high factuality users.

To assess the robustness of our results, we tested the consistency of the observed effects across eight alternative dataset constructions, varying both the user filtering criteria and the factuality thresholding. We found that tweet frequency and days since registration consistently distinguished between high and low factuality users across all variations, with stable effect directions and magnitudes. Follower count also showed a qualitatively consistent pattern, with a stable difference between high and low factuality groups, although the effect size varied somewhat across conditions. The number of followed accounts showed more variation, but the distinction between high and low factuality users—particularly the stronger negative association for high factuality users—was preserved across nearly all dataset versions. These results reinforce the general robustness of our findings while also highlighting that some metrics (e.g., follower count and followed account count) are more sensitive to dataset construction choices than others (see Figure 10 in the Supplementary Information).

Finally, we examined the interactions between the effects of the four social network metrics on factuality to unfold potential interplay of the independent variables. We have found that some of the social network metrics interact significantly with one another on factuality. To illustrate the interactions we divided the users into two equally sized groups based on the given social network metrics, e.g. followed account count, and plotted the effects at the median of each group.

\begin{figure}
      \centering
      \includegraphics[width=.95\textwidth]{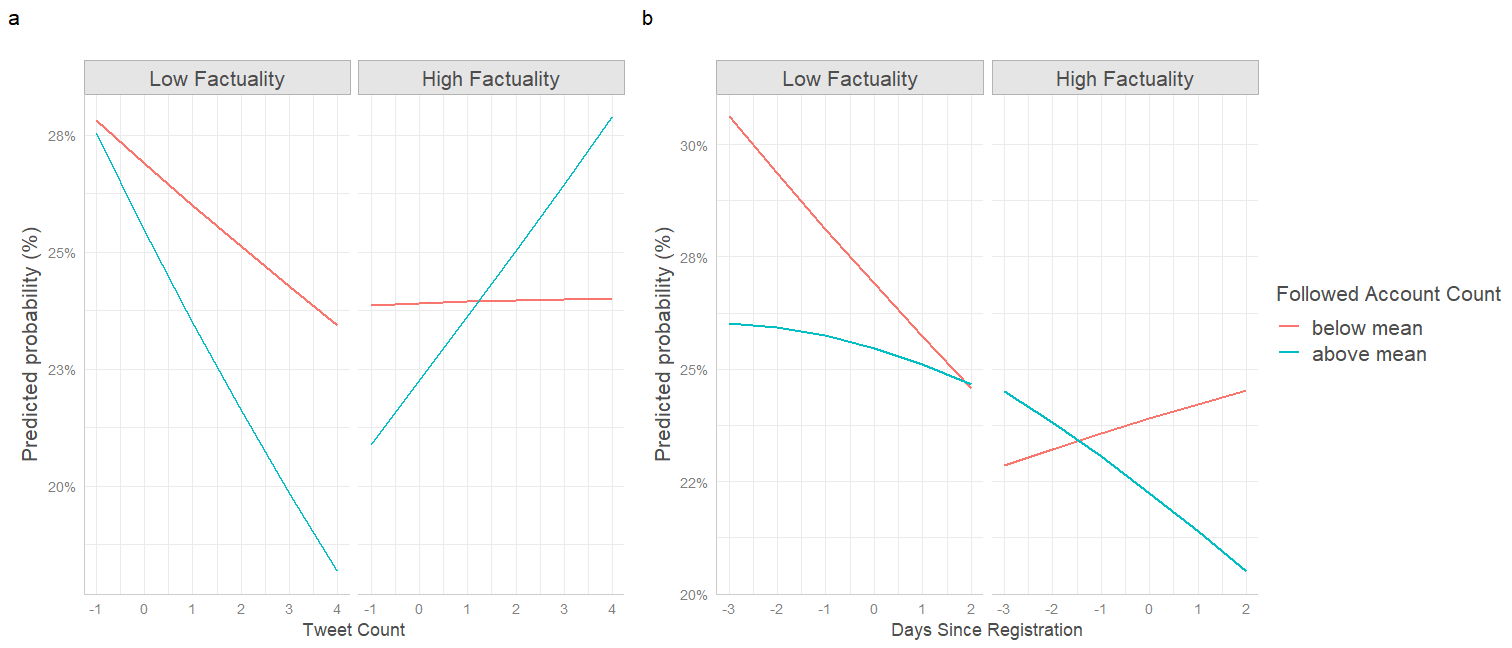}
      \caption{\textbf{Average marginal effect interactions between followed account count, average daily tweet count, and days since registration.} 
      Panel a: The effect of tweet activity on factuality differs by followed account count, with positive effects on high factuality and negative effects on low factuality primarily among those who follow many accounts.
      Panel b: The effect of days since registration on factuality also varies by followed account count, with opposite trends for high factuality and consistent negative association with low factuality.}

      \label{fig:fig6}
    \end{figure}

Our analysis reveals differences regarding the effect of the average daily tweet count on factuality (\hyperref[fig:fig6]{Figure 6}, panel a). In examining average daily tweet count across different levels of followed account count and its association with factuality levels we find that among those with few accounts followed the average tweet count does not affect the likelihood of belonging to the high factuality group. However, an increase in tweet activity is associated with an increased likelihood of belonging to the high factuality group among those who follow more accounts. For the low factuality group, tweet count is negatively associated with low factuality and this effect is stronger among those who follow more accounts. There is no significant difference in the likelihood of belonging to the low factuality group among those who tweet infrequently, however, among those who tweet frequently, users following few accounts are way more likely to have low factuality than users who follow more accounts.

In relation to days since registration, our analysis (\hyperref[fig:fig6]{Figure 6}, panel b) shows that among users who follow fewer accounts, those who have been registered on the platform for a longer duration are more likely to be in the high factuality category, while among those who follow more accounts, the trend is the opposite: the longer they are on the platform, the less likely that they belong to the high factuality group. The relationship between days since registration and low factuality does not vary based on followed account count. The relationship is consistently negative between days since registration and low factuality, indicating that longer registration periods are associated with a decreased likelihood of low factuality, however, this effect is somewhat more pronounced among users who follow fewer accounts.

In examining the interactions of the effects among the rest of the explanatory variables we found only slight differences. In only one of the other cases was there a difference in the direction of the effect based on another variable (effects of followed account on high factuality by follower count). The rest of the observed differences are small, regarding either the magnitude or the strength of the effects (see the full set of explanatory variable interactions in Supplementary Figure 11).

\section*{Discussion}
In this study, we aimed to examine whether simple, low-barrier metrics are systematically associated with differences in users’ likelihood of sharing misinformation on the social media X (Twitter). Drawing upon existing literature and leveraging easily accessible data from X, we tested four social network metrics as possible correlates of user factuality. Our analysis focused on follower count, average daily tweet count, number of users followed, and account age. The results offer several insights into the effectiveness of these metrics as group-level indicators of misinformation sharing tendencies.

Our findings indicate that when considered individually, each of the four metrics—follower count, average daily tweet count, number of users followed, and account age—has a significant relationship with user factuality scores. However, when all metrics were analysed together using multinomial regression, the distinctive power of follower count and followed account count diminished for the separation of low and high factuality, while average daily tweet count and number of users followed remained robust correlates of low and high factuality scores.

Higher average daily tweet counts were linked to lower factuality scores, consistent with our outlined expectation grounded in previous research which suggests that the sheer volume of content users engage with can detract from their ability to discern true from false information. This may be because the higher the number of tweets one shares, the less attention each individual piece of content would receive, making the user more likely to share non-accurate content \cite{Apuke2022}.

Older accounts were associated with higher factuality scores, in line with our expectations that longer presence on the platform correlates with higher content accuracy, which may be moderated by better digital literacy. The finding might be explained by the fact that users with older accounts might have developed stronger digital literacy (specifically on X) as they have had more time to do so, whereas this is not yet true for users with newer accounts, in line with findings from \cite{Guess2020}. Furthermore, according to diffusion of innovations theory \cite{Rogers2003}, those who adopt new technologies earlier are often more digitally literate, which could explain why older accounts are more factual. We might also find fewer old accounts with low factuality because of certain policies of X removing strongly non-factual accounts.

Finally, the multinomial regression analysis revealed certain significant interaction effects in the case of follower and followed account count. Among users who follow few accounts, tweeting more is less strongly associated with low factuality than among those who follow many accounts, while the effects on high factuality are positive for users who follow many accounts. Among users who follow many accounts, account age is less strongly associated with low factuality than among those who follow few accounts, while among users who follow few accounts, longer time on the platform is linked to higher factuality, whereas among those who follow many accounts, longer platform use is associated with lower factuality. These cross-effects highlight the context-dependent nature of the relationship between these social network metrics and factuality. 

These findings have several important implications. For social media platforms, the results suggest that basic metrics like tweet count and account age could inform efforts to understand which types of users are more likely to share misinformation on average, allowing for targeted interventions such as accuracy prompts or digital literacy campaigns. Policymakers could advocate for the incorporation of these metrics into misinformation mitigation strategies, ensuring a low-cost, scalable approach. Future research should continue to explore additional accessible data points and refine existing metrics to better understand the behavioral tendencies associated with misinformation sharing.

While our approach provides valuable insights into the identification of users likely to share misinformation on X, several limitations must be acknowledged. First, it is important to note that the analysis presented here is  not aimed at individual-level prediction. Instead, we aim to identify broad trends in user behavior that may be associated with misinformation sharing, rather than develop a robust predictive model. As such, the overlap in distributions among high- and low-factuality users limits the ability to draw definitive conclusions about individual users’ likelihood of spreading misinformation.

Moreover, the dataset we used originates from a specific population—followers of a sample of users recruited via Prolific who self-reported their X handles in a survey on vaccine hesitancy. While this initial sample may introduce some bias, our analyses were not limited to the survey participants themselves but expanded to include their first-degree followers. This significantly larger group—over 11,000 users—likely offers a broader behavioral snapshot of the general Twitter population, though some residual similarities may remain.
However, rather than aiming for full population representativity, the purpose of our dataset is to compare the behavior of users who are more prone to share content from unreliable sources to those who are more likely to share reliable ones.

Importantly, our decision to use this dataset was guided by the goal of studying regular users. Many available public datasets that support factuality scoring rely on keyword-based data collection, which tends to over-represent highly active or politically engaged accounts. These are often not reflective of typical user behavior. In contrast, our dataset—filtered to exclude bots, verified users, and extreme activity levels—better represents everyday users and their misinformation-related behaviors. As such, while not perfectly representative, it provides an appropriate test case for our research goals.

Beyond sampling considerations, another limitation of our method is the exclusion of users who share domains without Media Bias/Fact Check (MBFC) ratings. This decision restricts the scope of our analysis, as it may not be fully representative of the broader user base, particularly those sharing content from sources that are not evaluated by MBFC. We collected the latest 500 tweets containing URLs for approximately 1.6 million users. Out of these, 240,915 users had ever shared a URL, and among them, only 59,610 had shared at least one URL that could be classified using MBFC ratings. We report the proportion of such users excluded during our filtering process in the Methods section to provide transparency on the scale of this limitation. Future research could explore the implications of including such users to better understand the full spectrum of misinformation-sharing behavior.

Beyond limitations related to sampling and content coverage, the generalizability of our findings is influenced by the platform-specific nature of user behavior. X, a highly public, text-based platform, is distinct from other social media platforms such as Facebook, which emphasizes private networks and long-form content. The social network metrics we use to identify misinformation spreaders on X may not be as effective on platforms where user engagement patterns differ significantly. Therefore, while our findings offer insights specific to X, they may not fully translate to other platforms with different interaction styles.

Network structure variations between platforms also present challenges. X operates on a directed follower-following model, where information flows asymmetrically, whereas platforms such as Facebook and LinkedIn use mutual connections. This structural difference may result in variations in the usefulness of network-based features like follower count and following count when applied to other platforms with more reciprocal interactions.

The impact of content moderation policies further complicates the application of our approach across platforms. Each platform has unique strategies for content distribution and moderation—such as visibility algorithms, labeling practices, and content removal—that shape the dynamics of misinformation spread. These differences may affect how observable and measurable misinformation sharing is, and thus influence the performance of detection methods like ours.

Despite these limitations, our methodological framework, which leverages social network metrics for early identification of misinformation spreaders, holds promise for adaptation to other platforms. However, careful consideration of platform-specific differences is necessary, and future work should explore these variations to determine whether similar patterns of behavior emerge across different social media environments.

In conclusion, this study demonstrates that basic, easily accessible social network metrics, particularly tweet count and account age, can help characterize differences between low and high factuality users. While the studied metrics may offer a scalable and efficient way to explore behavioral patterns associated with groups of users more likely to share misinformation, they should be interpreted with caution and in context. Their use may be particularly helpful for informing broader strategies aimed at understanding or addressing misinformation dynamics, offering a foundation for cost-effective interventions to curb the spread of misinformation on social media platforms. However, their usefulness may vary across platforms due to differences in user behavior, content moderation policies, and network structures.

\section*{Methods}
    \subsection*{Data}
    Our study builds upon the data used in the studies by Rathje et al. \cite{Rathje2022}, which was approved by the University of Cambridge Psychology Research Ethics Committee (PRE.2020.144). All participants in the study provided informed consent before answering surveys and providing their Twitter usernames~\cite{Rathje2022}.
    This dataset was selected in part due to its unique ability to capture regular, human user behavior at scale. Most public misinformation datasets rely on keyword-based collection methods that tend to over-represent highly active or partisan users. Our focus on followers of original survey participants allowed us to build a larger, more behaviorally representative dataset for this purpose.

    
    The study collected 463 survey responses through the online recruitment platform Prolific. Starting from those usernames, the publicly accessible followers of those accounts were collected, resulting in a dataset of n = 1,670,127 users.
    Such data includes information about the number of followers and following, as well as the number of content posted and the subscription date. For each user, the latest 500 tweets containing a URL to external sources, posted within one year of the most recent published content, were collected.
    In this study, we utilize users' data, including the number of followers of the account, the number of people followed by the account, the total number of tweets produced by the account, registration dates, and their tweets containing a URL, all directly extracted using the official Twitter API.
    All methods in this study were performed in accordance with the relevant guidelines and regulations.
    
    \subsection*{User factuality}
    The reliability of users' diets was measured using a metric called Factuality, based on the domains of the content they published.
    The quantification of this metric relies on data from Media Bias Fact Check (MBFC), an independent fact-checking organisation that classifies news outlets based on both their reliability and political bias \cite{MediaBiasFactCheck2023}. Each news outlet on MBFC is assigned labels indicating its reliability (Very Low, Low, Mixed, Mostly Factual, High, Very High). 
    News outlets’ factuality corresponds to MBFC's reliability label. These metrics have been applied individually to each X user (denoted as u), generating a vector, $F_u \in \mathbb{R}^{6\times1}$, representing the percentage of links shared by the user for Factuality categories.
    We assign each user a unique factuality score, calculated as the average of the factuality scores associated with their posted tweets.
    To assign factuality scores, we initially collected up to the latest 500 tweets containing URLs, posted within one year of the most recent published content, for the approximately 1.6 million users. Out of these, only 240,915 users had ever shared a URL, and among them, only 59,610 had shared at least one domain we were able to classify using MBFC ratings.
    We segmented users into low, middle, and high factuality groups based on their respective factuality scores. This involved dividing users into the three factuality categories such that the bottom and top 25\%, 30\%, or 35\% would be the low and high factuality users, respectively, and the remaining middle would be the middle factuality group, resulting in three distinct factuality groups. Our primary goal with this approach is to identify the most prominent misinformation spreaders and characterize their behaviors, rather than provide a continuous score for all users. While continuous scores may offer more granularity, prior work has shown that misinformation sharing on social media follows a highly skewed distribution, with a small fraction of users responsible for the majority of spread \cite{Grinberg2019}. Therefore, segmenting users allows us to better focus on actionable insights relevant to platform moderation and intervention strategies.
    Although our segmentation approach may appear to contrast with literature highlighting power-law distributions in misinformation sharing, it serves a different purpose: to provide a broad behavioral overview rather than pinpoint only extreme cases. Grouping users in the bottom 30\% (or 25\%, 35\%) allows us to capture general patterns among users with lower factuality.
    While the main focus of our paper lies on the results derived from the factuality categorisation taking the top and bottom 30\% as low and high factuality users, respectively, we also present findings obtained applying the other two thresholds in the Supplementary Information to validate the robustness of our conclusions.

    \subsection*{Filtering the dataset for regular users}
    As the purpose of our study is understanding the behaviour of “human users” \cite{Varol2017}, we initially curated our datasets to focus on users who are more representative of organic human social media users. I.e., we wanted to exclude accounts from the study that appeared to be either business, political or celebrity accounts, or bots. To ensure the selection of regular individuals, we implemented specific criteria as outlined below. These criteria were derived from various sources and were applied iteratively to curate our dataset.\\
    Verified status: Verified accounts, typically belonging to recognized individuals or brands, were identified and removed from the dataset as they do not represent typical user behaviour. This was based on X's verification policy—a rigorous review process of the authenticity, notability, and activity of the account at the time of the data collection \cite{TwitterHelpCenter}. Such accounts comprise 2.9\% of all accounts in the dataset.\\
    Average number of tweets per day: Accounts with an exceptionally high average number of tweets per day were considered unlikely to belong to regular users, as this behaviour is more characteristic of commercial or political accounts, sometimes bots \cite{PewResearchCenter2019}. We set thresholds to remove accounts with tweet/day ratios exceeding certain values expected to be limits of regular user behaviour.\\
    Number of followers: Accounts with a very high number of followers were considered atypical for regular users, as such levels of fame are usually attained outside of X or through exceptionally high activity on the platform. We removed accounts with follower counts surpassing a certain threshold.\\
    Follower/followed account ratio: A high proportion of followers compared to followed accounts is characteristic of celebrity-like accounts rather than typical users. Therefore, we removed accounts with follower/followed account ratios exceeding certain values.\\
    Number of followed accounts: Accounts with an unusually high number of followed accounts may indicate bot-like behaviour, particularly if the followings are not reciprocal \cite{Gilani2017}. We set thresholds to remove accounts with followed account counts exceeding specific values.\\
    Followed account/follower Ratio: A high proportion of non-reciprocal followings may also indicate bot-like behaviour \cite{Gilani2017}. Therefore, accounts with followed account/follower ratios exceeding certain values were removed.
    Finally, to filter for bots, we applied an additional filtering step using Botometer \cite{davis2016botornot}, specifically removing users with a Botometer score above 0.4, following the “human threshold” suggested by \cite{Pozzana2020}.
    
    Based on the stringency of the above introduced criteria we created three distinct groups of datasets: relaxed, middle, and strict. By applying these criteria, we aimed to ensure the selection of regular individuals whose behaviour is more representative of typical social media users for our analysis. The specific values for each strictness category are shown in Table \ref{table:table1}.

    To characterize the groups resulting from our filtering criteria, we present the distribution of Factuality scores in Figure 1 in the Supplementary Information, illustrating the differences between the selected groups.

    In addition to providing the distributions, we briefly characterize the groups based on their factuality scores. As shown in Supplementary Figure 1, low factuality users are predominantly characterized by scores below 60, middle factuality users span scores between 60 and 70, and high factuality users are characterized by scores above 70, with a strong concentration around 80. We note that the observed peaks in the distributions arise from the underlying labeling procedure: factuality scores are assigned based on categorical source credibility ratings, leading to discrete jumps at 0, 20, 60, 80, and 100. As many users' sharing histories contain sources with identical credibility ratings, sharp spikes are observed in the plotted distributions at scores 40, 60, and 80.

    \begin{table}[h!]
        \centering
        \begin{tabular}{|>{\bfseries}l|>{\centering\arraybackslash}p{3cm}|>{\centering\arraybackslash}p{3cm}|>{\centering\arraybackslash}p{3cm}|}
        \hline
        \multicolumn{1}{|c|}{\textbf{Criteria}} & \textbf{Relaxed} & \textbf{Middle} & \textbf{Strict} \\ \hline
        Verified Status            & -         & Not Verified  & Not Verified \\ \hline
        Average Tweets Per Day     & $\leq 56$ & $\leq 32$     & $\leq 16$ \\ \hline
        Follower Count             & $\leq 10,000$ & $\leq 10,000$ & $\leq 5,000$ \\ \hline
        Follower/Followed Account Ratio   & $\leq 10$ & $\leq 5$      & $\leq 3$ \\ \hline
        Followed Account Count            & $\leq 10,000$ & $\leq 10,000$ & $\leq 5,000$ \\ \hline
        Followed Account/Follower Ratio   & $\leq 10$ & $\leq 10$     & $\leq 5$ \\ \hline
        Botometer score   & $\leq 0.4$ & $\leq 0.4$     & $\leq 0.4$ \\ \hline
        \end{tabular}
        \caption{Criteria for dataset variation for robustness check.}
        \label{table:table1}
    \end{table}

    As in the case of factuality grouping, the main focus of our paper lies on the results derived from the middle category, but we also present findings obtained applying the two other criteria in the Supplementary Information to validate the robustness of our conclusions.
    The number of users in each dataset are shown in Table \ref{table:table2}.

 \begin{table}[h!]
    \centering
    \begin{tabular}{|>{\centering\arraybackslash}p{3cm}|>{\centering\arraybackslash}p{3cm}|>{\centering\arraybackslash}p{3cm}|>{\centering\arraybackslash}p{3cm}|}
    \hline
    \multicolumn{1}{|c|}{\textbf{Criteria}} & \textbf{Relaxed} & \textbf{Middle} & \textbf{Strict} \\ \hline
    \textbf{Top/bottom 25\%} & 4492 & 4218 & 2878 \\ \hline
    \textbf{Top/bottom 30\%} & 5427 & 5113 & 3643 \\ \hline
    \textbf{Top/bottom 35\%} & 6166 & 5806 & 4072 \\ \hline
    \textbf{Total} & 17616 & 16586 & 11503 \\ \hline
    \end{tabular}
    \caption{Number of observations in each of the datasets used for validation of results.}
    \label{table:table2}
\end{table}

    \subsection*{Randomised dataset}
    To ensure the conclusions drawn from our results are not due to chance, we constructed a counterfactual dataset consisting of 1,000 datasets mirroring the structure of our original dataset. In these counterfactual datasets, factuality scores were randomly shuffled across individuals while keeping all other characteristics constant.

    \subsection*{Analysis of combined effects}
    We employed multinomial regression analysis to investigate the combined effects of our four social network metrics—follower count, followed account count, tweets per day, and days since registration—on factuality, which is categorised into three levels: low, middle, and high. Multinomial regression is particularly suitable for this analysis as it allows for the modelling of outcomes with multiple discrete categories, enabling us to examine the influence of multiple predictors on the probability of an observation belonging to each factuality level simultaneously.
    To interpret the results, we utilised Average Marginal Effects (AMEs). AMEs offer a clear and intuitive measure of the impact of each social network metric on the probability of an account being classified into each factuality category. By translating regression coefficients into average changes in probabilities, AMEs enhance the interpretability of our findings, making it easier to understand and compare the effects of different metrics. This approach aligns with the recommendations by \cite{Mood2010}, who highlights the advantages of using marginal effects in complex regression models.

\bibliography{main}

\section*{Acknowledgements}
J.K. acknowledges funding from the Hungarian Academy of Sciences Lendület Program: LP2022-10/2022.

\section*{Author contributions statement}



All authors designed the study. A.G. collected and processed the data. J.S. carried out the data analysis and statistical modeling. E.O., J.K. and A. G. supervised the study. All authors reviewed and approved the final manuscript.

\section*{Additional information}

\textbf{Data availability.} The X (Twitter) data is made available in accordance with X’s terms of service. Tweet IDs for each COP are available at: [the link will be provided upon publication]. The corresponding tweets can be downloaded using the official Twitter API (https://developer.twitter.com/en/docs/twitter-api).\\
\textbf{Competing interests.} The authors declare no competing interests.

\newpage
\section*{Supplementary Information for ``Easy-access online social media metrics are associated with misinformation sharing activity''}

Júlia Számely, Alessandro Galeazzi, Júlia Koltai, and Elisa Omodei

\setcounter{figure}{0}

\subsection*{Robustness checks}

\subsubsection*{{Distributions of factuality scores}}

\begin{figure}[h]
\centering
\includegraphics[width=.95\textwidth]{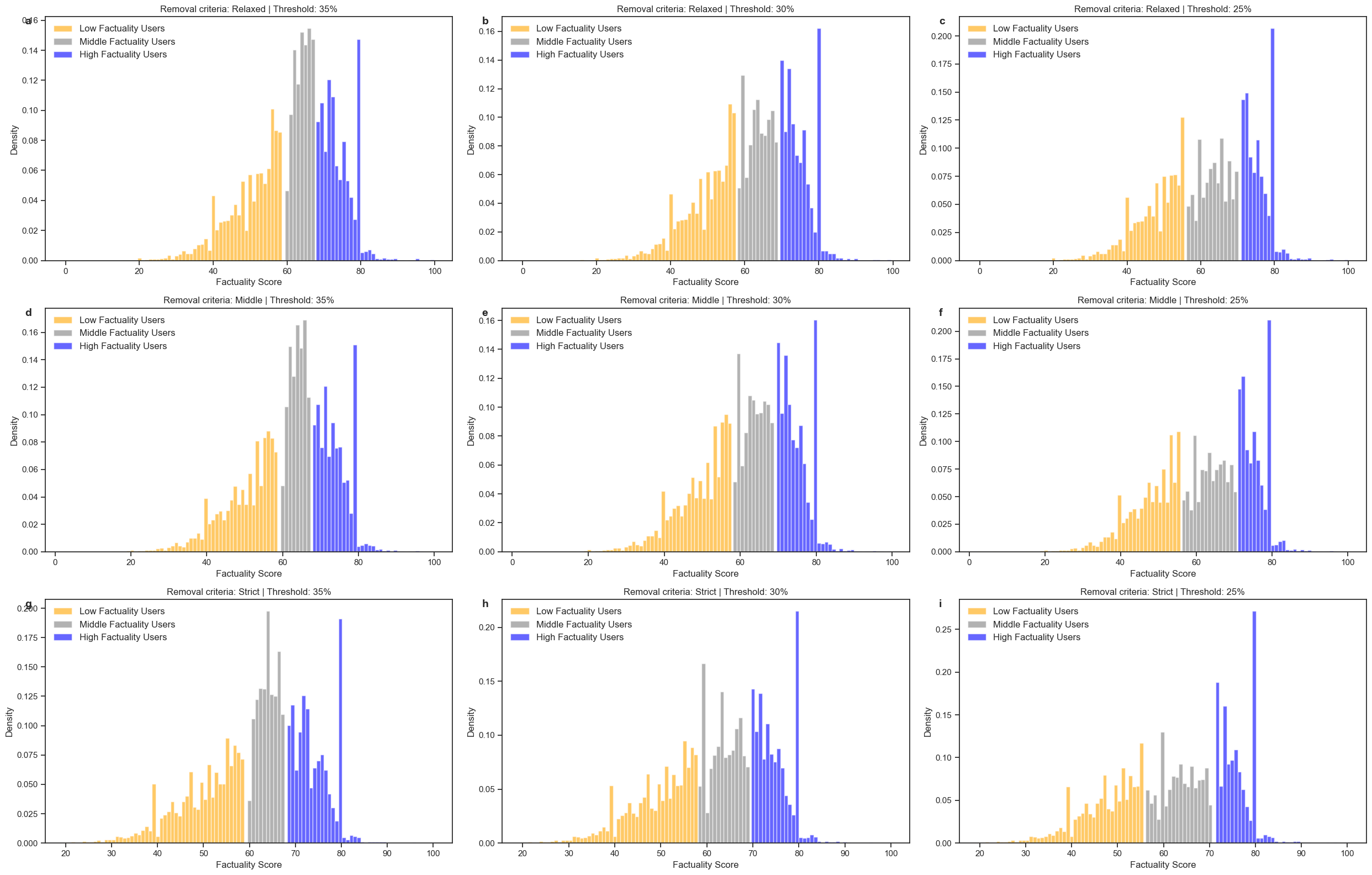}
\caption{\textbf{Distribution of factuality scores across the nine groups, defined by varying levels of regular user and bot filtering criteria (Relaxed, Middle, Strict) and low/high factuality group cutoffs (35/35, 30/30, 25/25).}}
\label{fig:supp-fig1}
\end{figure}

\clearpage
\subsubsection*{Distributions of follower count}

\begin{figure}[h]
\centering
\includegraphics[width=1\textwidth]{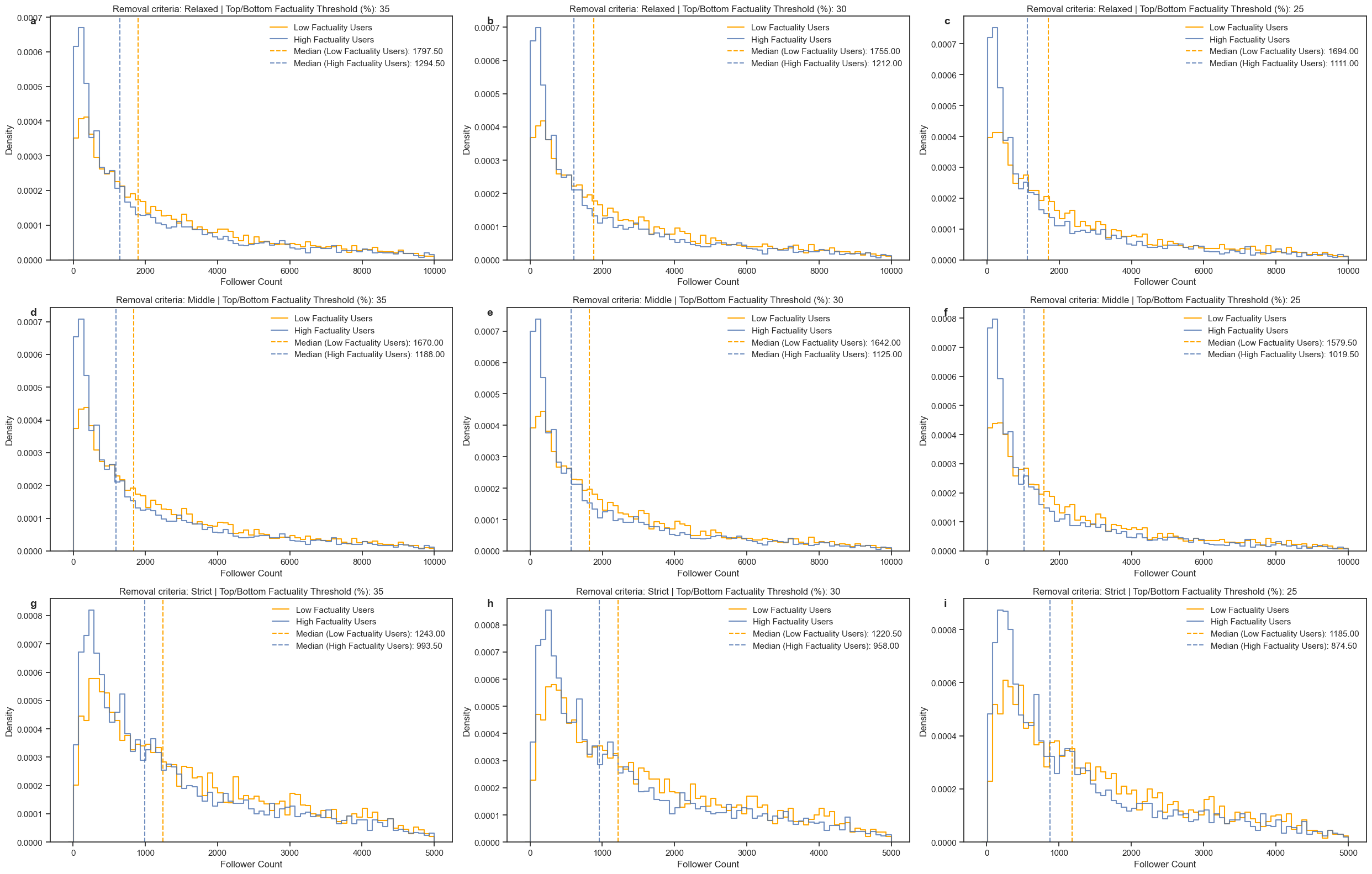}
\caption{\textbf{Low factuality users tend to have higher follower count than high factuality users.} Comparing the distributions of follower count between low (orange curve) and high factuality (blue curve) users, we find that low factuality users have a significantly higher number of followers compared to high factuality users (p-value $<$ 0.0001) in each of the tested datasets. The median value for low factuality users (orange dotted line) is higher than for high factuality users (blue dotted line) in each of the tested datasets.}
\end{figure}

\begin{figure}[h]
\centering
\includegraphics[width=1\textwidth]{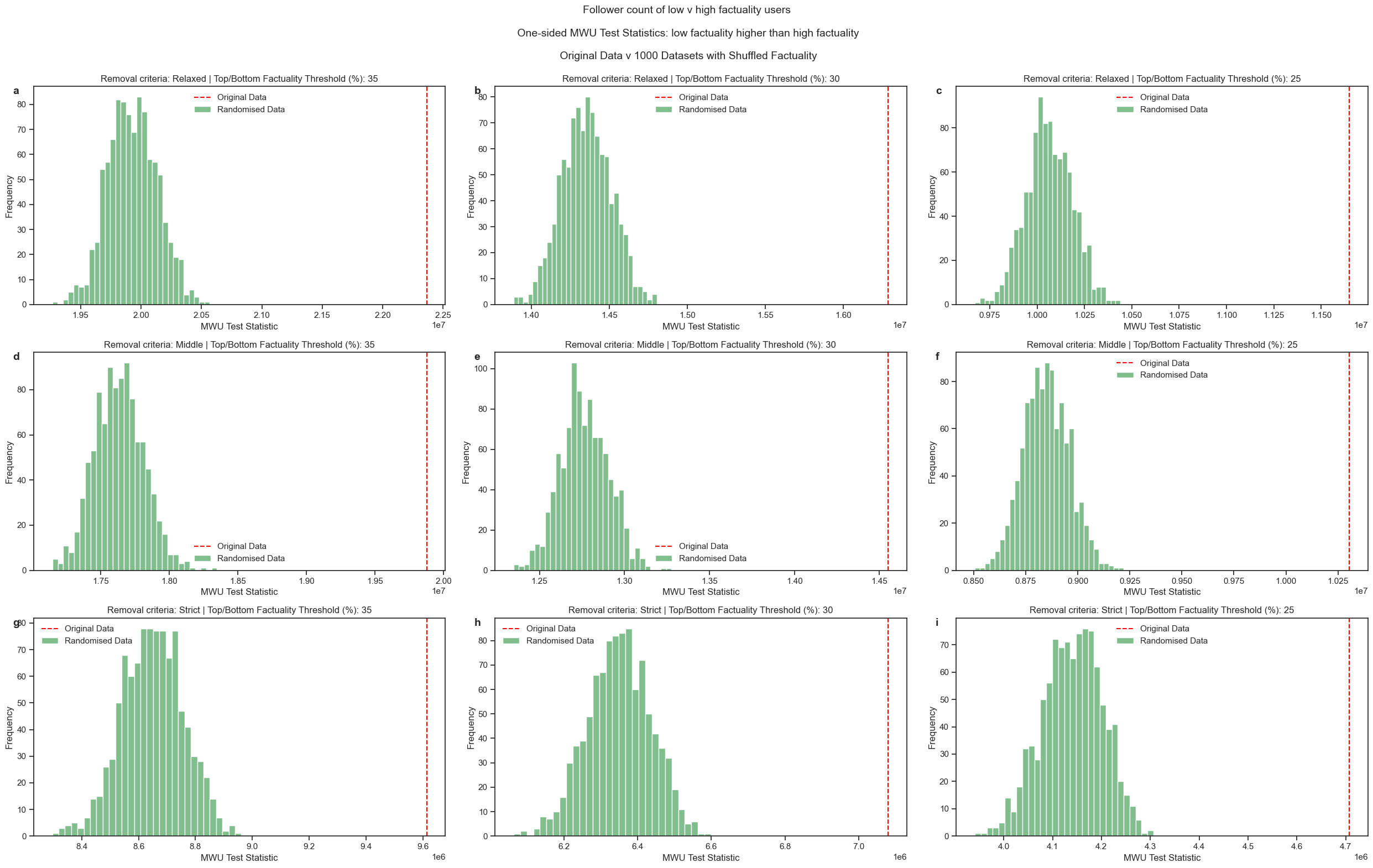}
\caption{\textbf{Low factuality users tend to have higher follower count than high factuality users.}    The MWU test score obtained from the empirical data (red dotted line) is higher than all the MWU test scores calculated on the 1,000 shuffled datasets (green bars) in each of the tested datasets.}
\end{figure}

\clearpage
\subsubsection*{Distributions of tweet count}

\begin{figure}[h]
\centering
\includegraphics[width=1\textwidth]{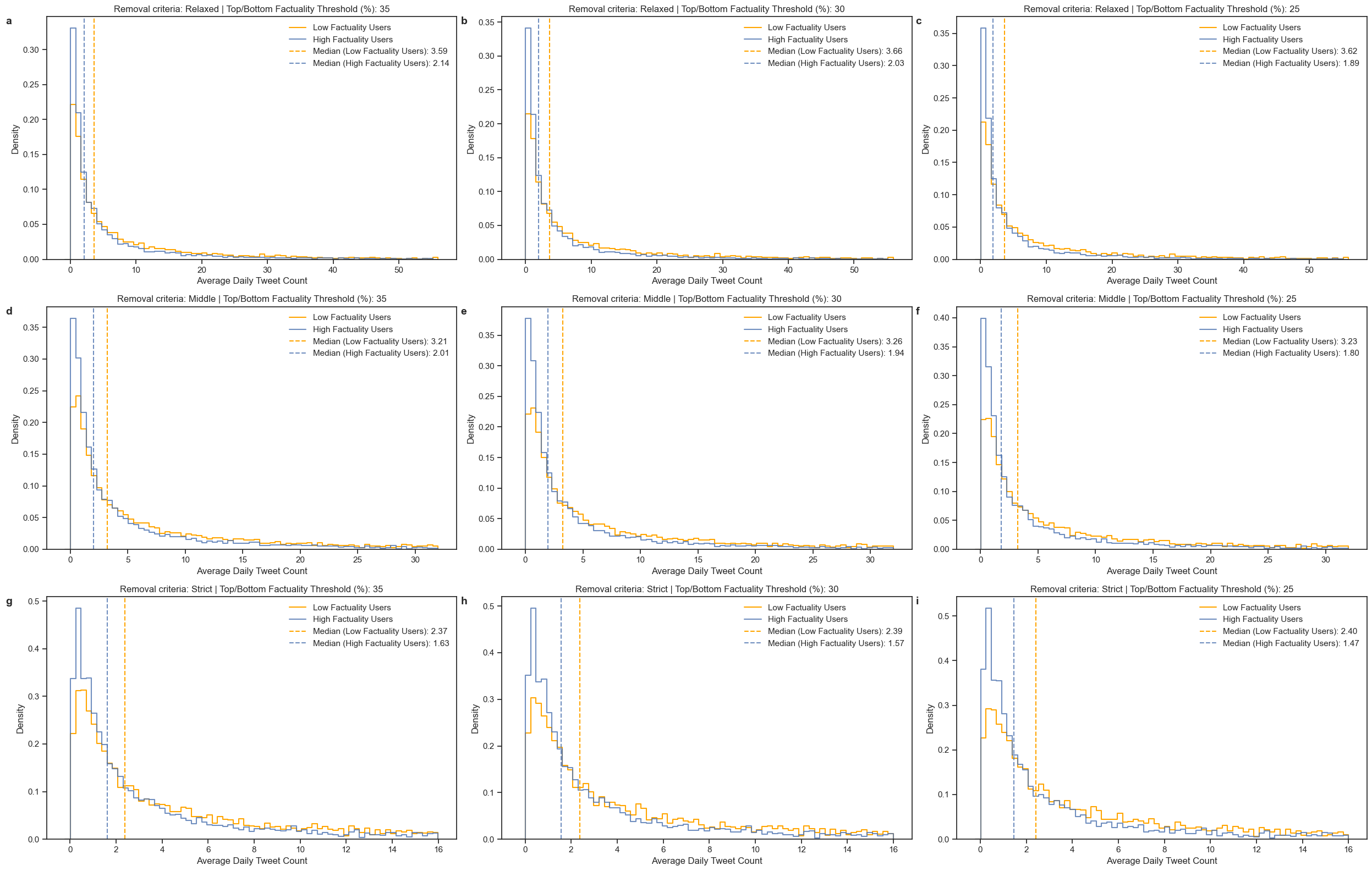}
\caption{\textbf{Low factuality users tend to have higher tweet count than high factuality users.} Comparing the distributions of tweet count between low (orange curve) and high factuality (blue curve) users, we find that low factuality users have a significantly higher number of tweets compared to high factuality users (p-value $<$ 0.0001) in each of the tested datasets. The median value for low factuality users (orange dotted line) is higher than for high factuality users (blue dotted line) in each of the tested datasets.}
\end{figure}

\begin{figure}[h]
\centering
\includegraphics[width=1\textwidth]{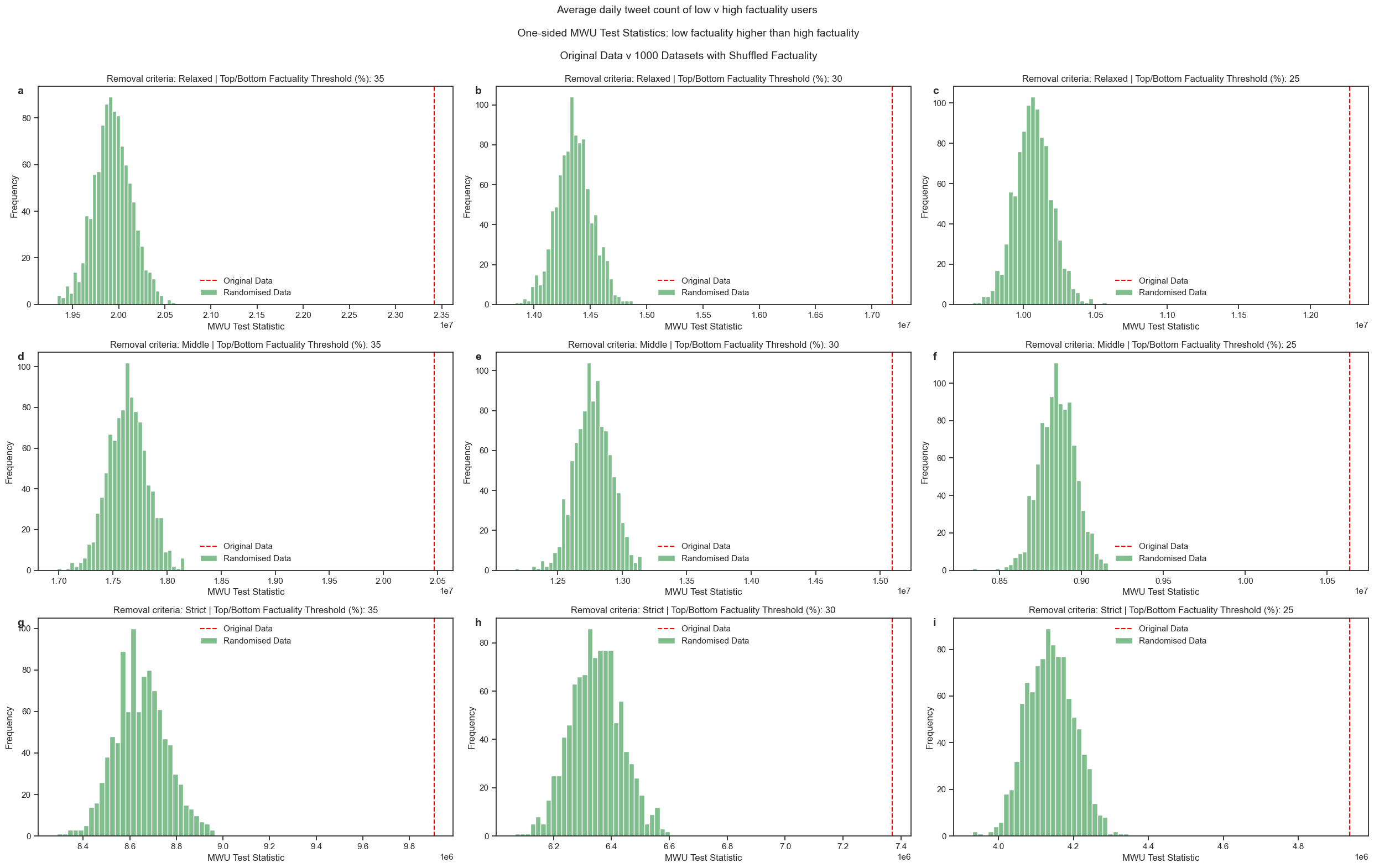}
\caption{\textbf{Low factuality users tend to have higher tweet count than high factuality users.} Panel b: The MWU test score obtained from the empirical data (red dotted line) is higher than all the MWU test scores calculated on the 1,000 shuffled datasets (green bars) in each of the tested datasets.}
\end{figure}

\clearpage
\subsubsection*{Distributions of followed account count}

\begin{figure}[h]
\centering
\includegraphics[width=1\textwidth]{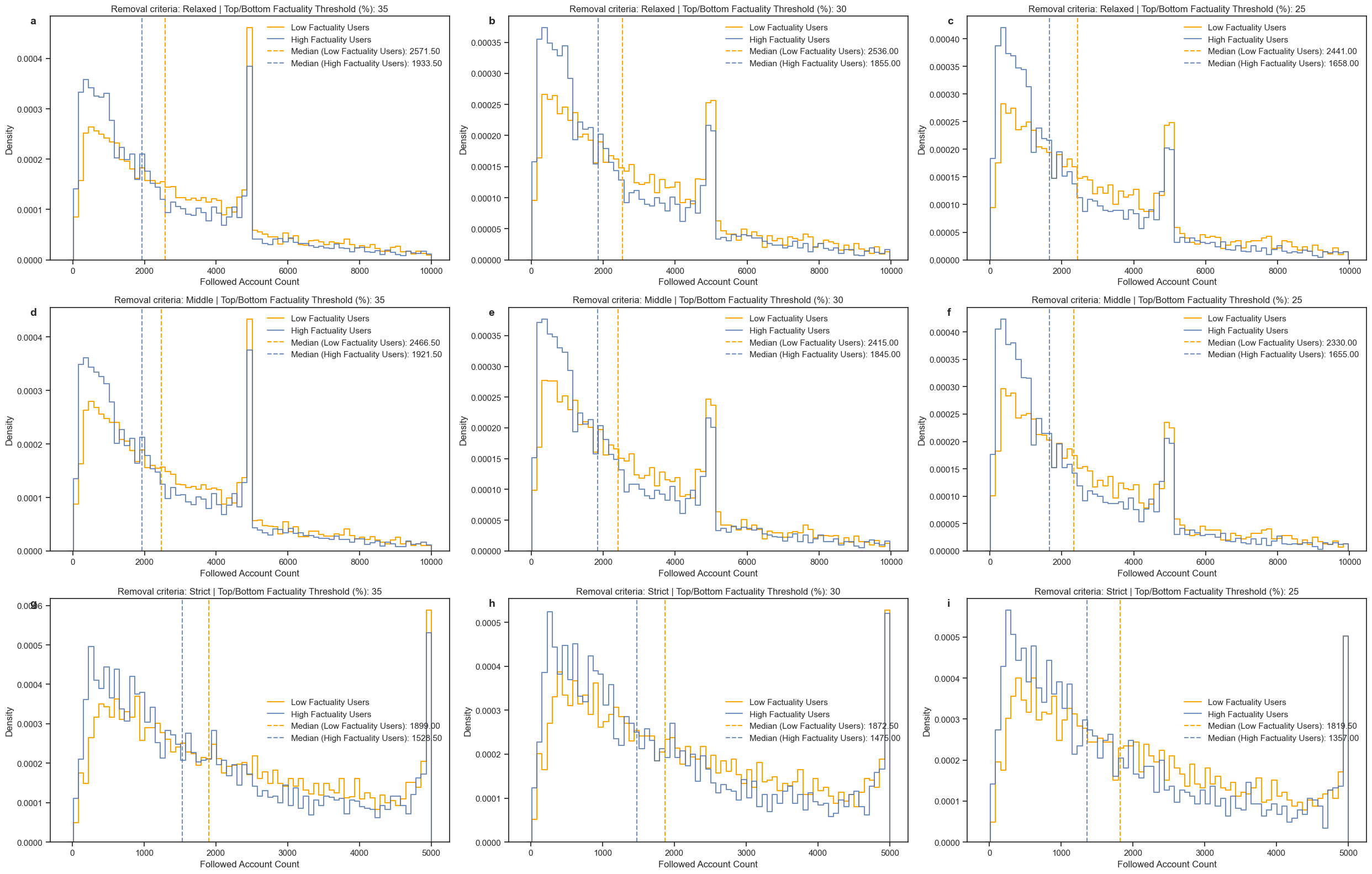}
\caption{\textbf{Low factuality users tend to have higher followed account count than high factuality users.} Comparing the distributions of followed account count between low (orange curve) and high factuality (blue curve) users, we find that low factuality users have a significantly higher number of followed accounts compared to high factuality users (p-value $<$ 0.0001) in each of the tested datasets. The median value for low factuality users (orange dotted line) is higher than for high factuality users (blue dotted line) in each of the tested datasets. The peak around 5000 is due to an X policy that limits the number of new followed accounts until the user obtains more followers.}
\end{figure}

\begin{figure}[h]
\centering
\includegraphics[width=1\textwidth]{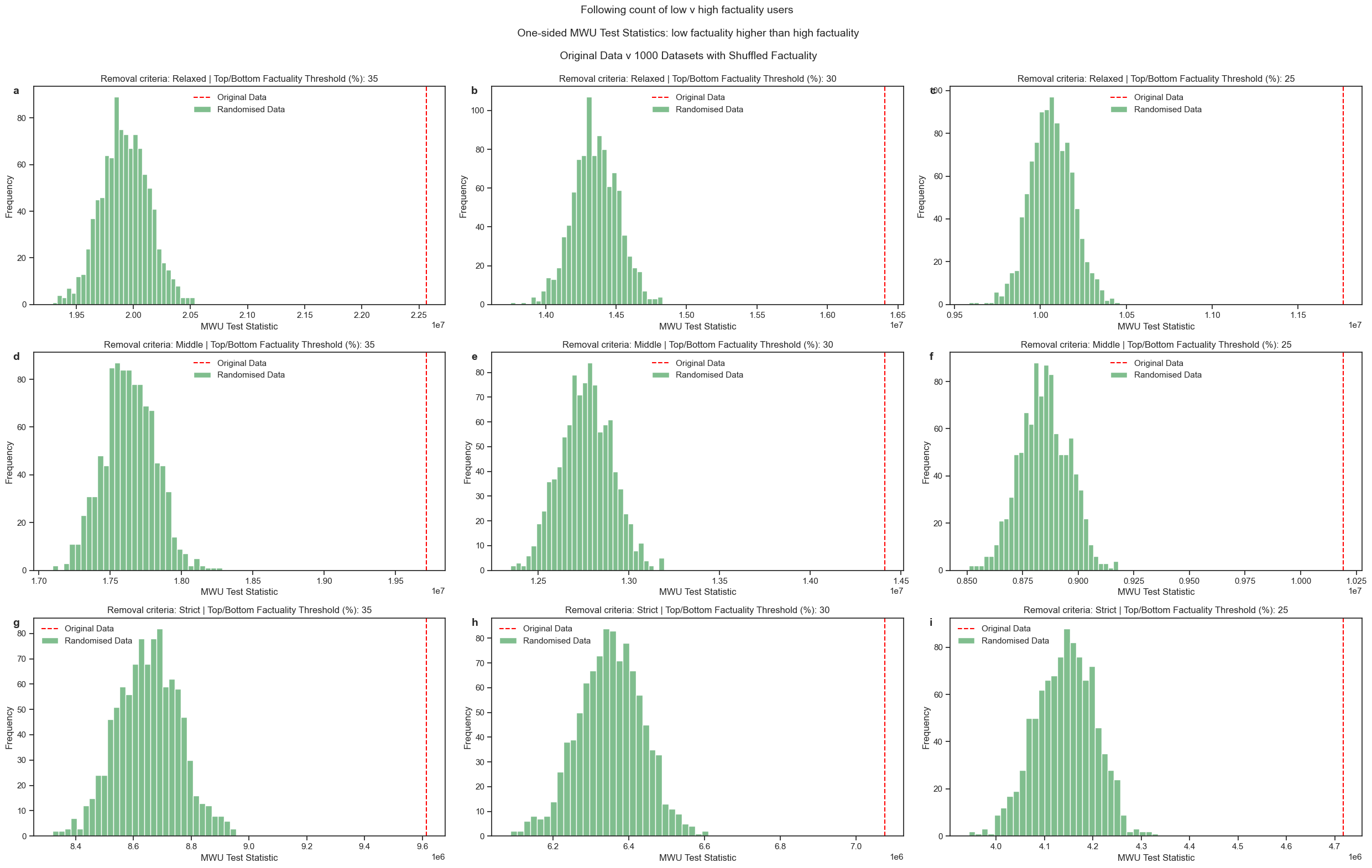}
\caption{\textbf{Low factuality users tend to have higher followed account count than high factuality users.} The MWU test score obtained from the empirical data (red dotted line) is higher than all the MWU test scores calculated on the 1,000 shuffled datasets (green bars) in each of the tested datasets.}
\end{figure}

\clearpage
\subsubsection*{Distributions of account age}

\begin{figure}[h]
\centering
\includegraphics[width=1\textwidth]{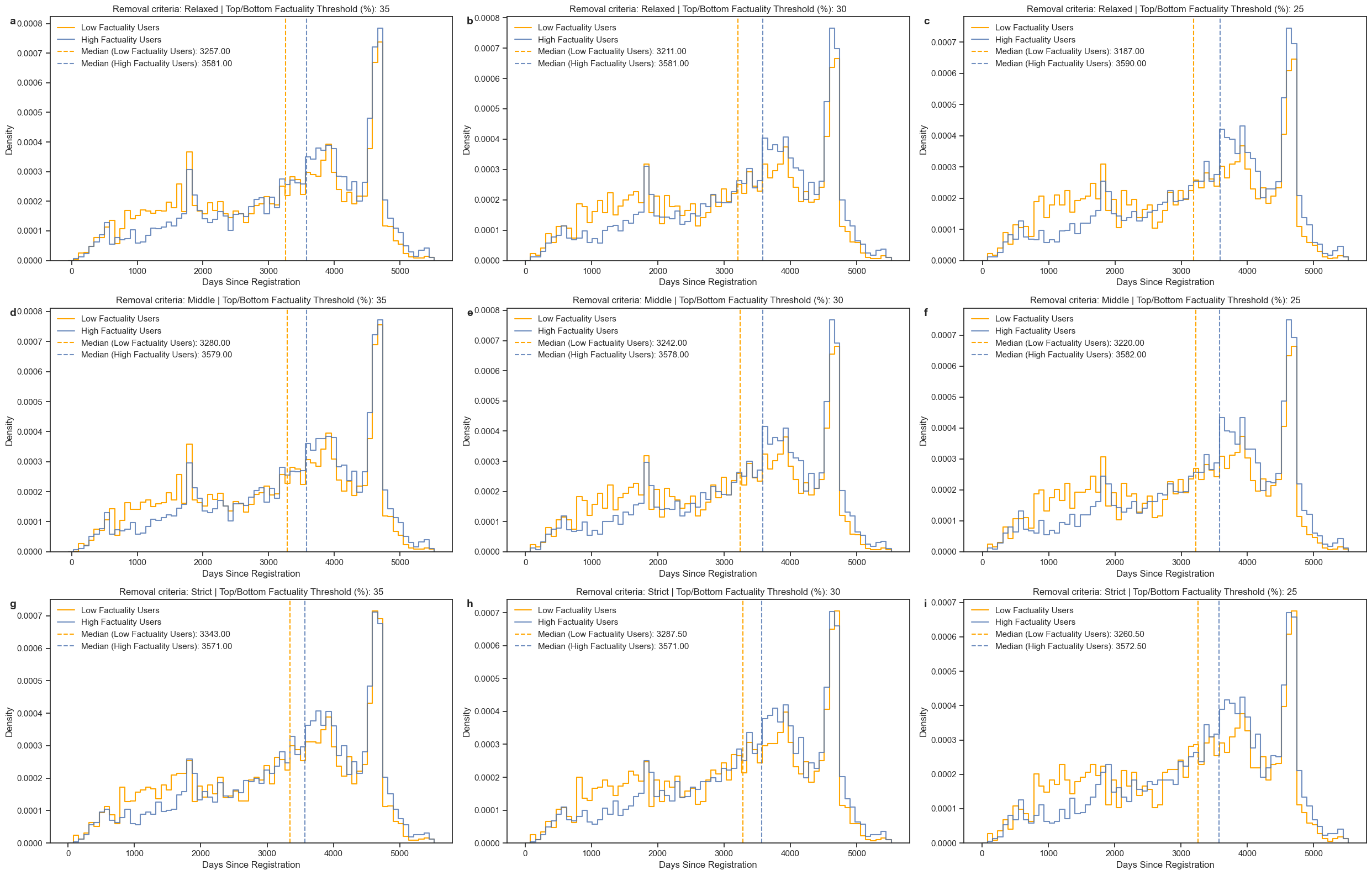}
\caption{\textbf{Low factuality users tend to have lower number of days since registration than high factuality users.} Comparing the distributions of the number of days since registration between low (orange curve) and high factuality (blue curve) users, we find that low factuality users have a significantly lower number of days compared to high factuality users (p-value $<$ 0.0001) in each of the tested datasets. The median value for low factuality users (orange dotted line) is lower than for high factuality users (blue dotted line) in each of the tested datasets.}
\end{figure}

\begin{figure}[h]
\centering
\includegraphics[width=1\textwidth]{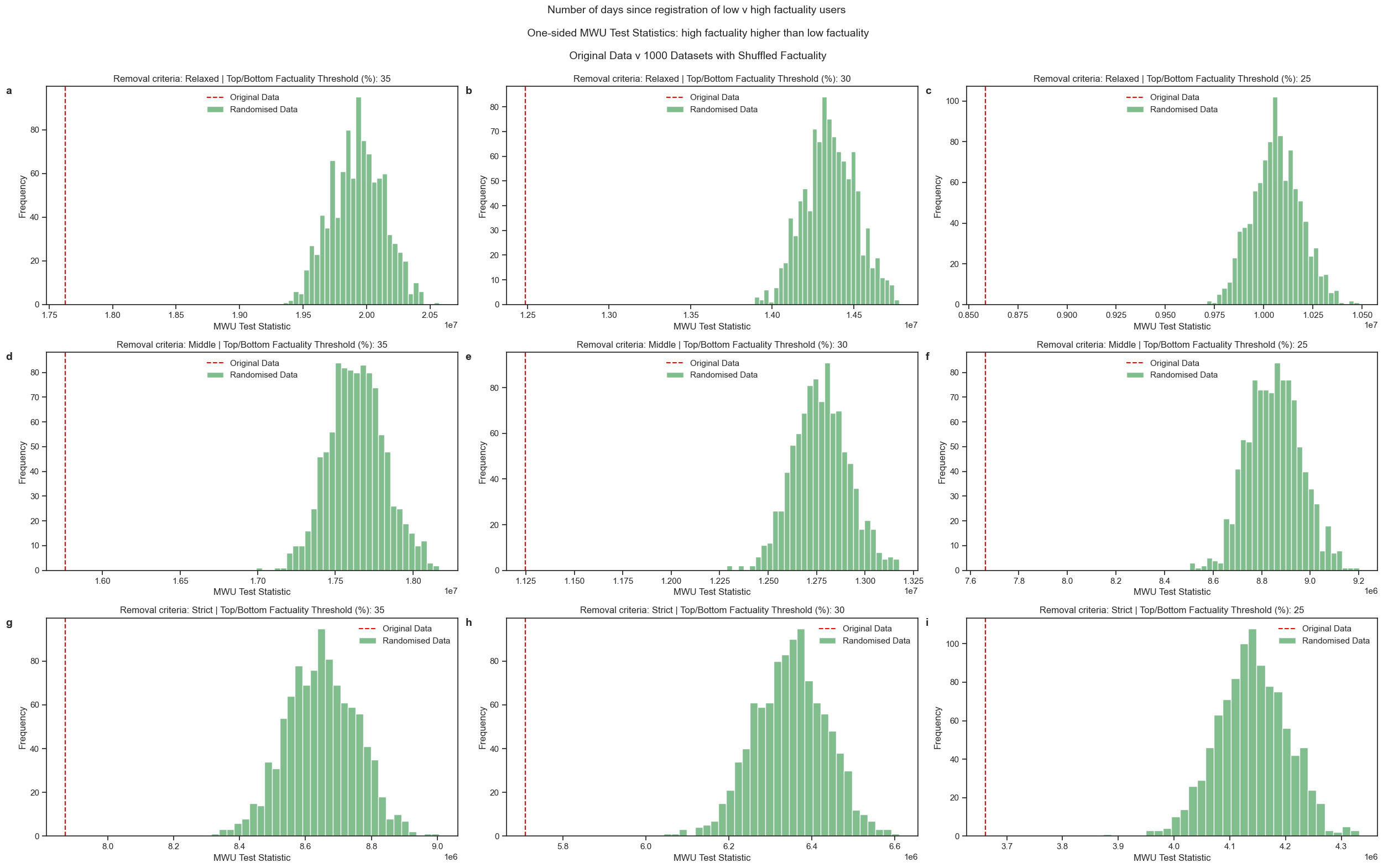}
\caption{\textbf{Low factuality users tend to have lower number of days since registration than high factuality users.} The MWU test score obtained from the empirical data (red dotted line) is lower than all the MWU test scores calculated on the 1,000 shuffled datasets (green bars) in each of the tested datasets.}
\end{figure}

\clearpage
\subsubsection*{Regression results}

\begin{figure}[h]
\centering
\includegraphics[width=1\textwidth]{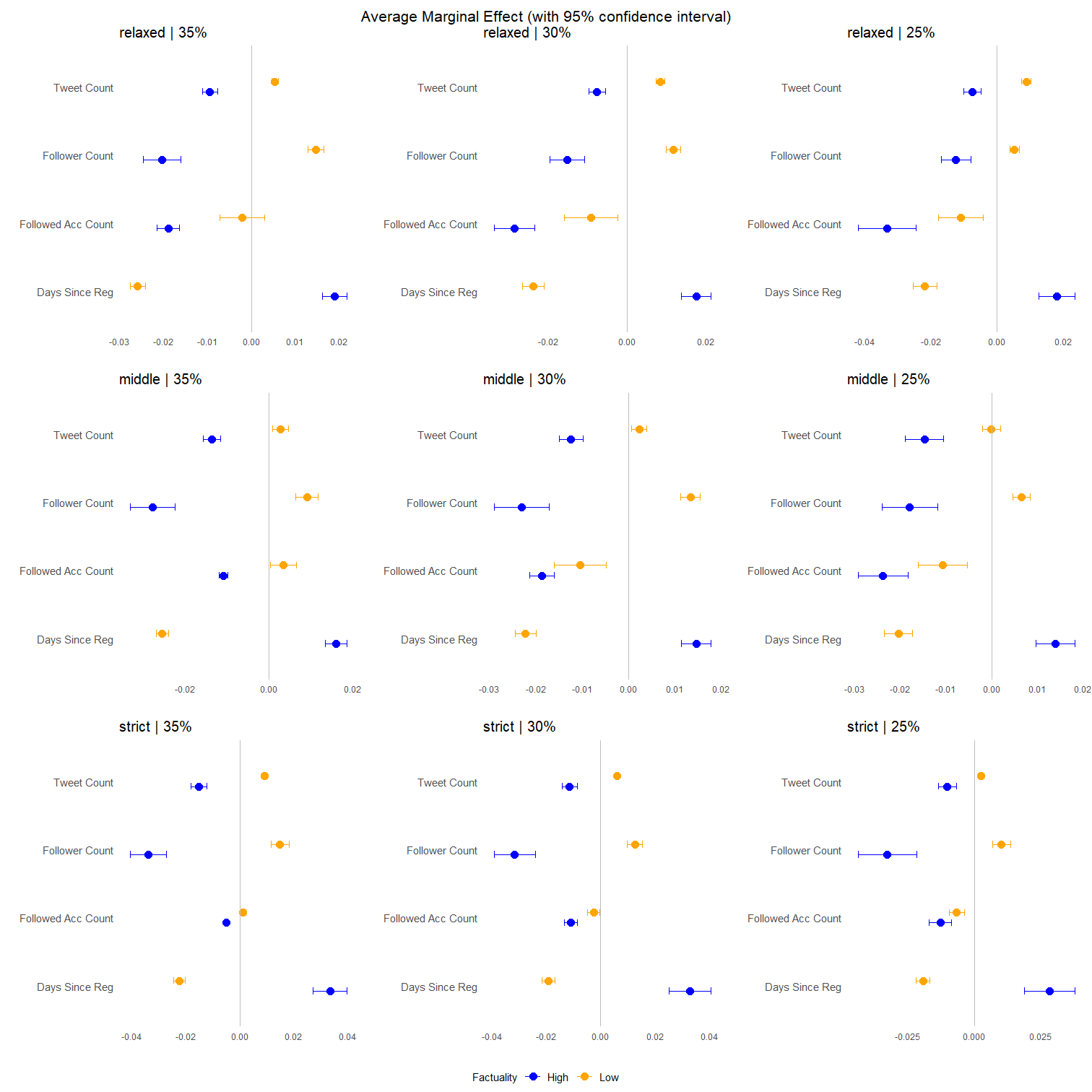}
\caption{Average marginal effects of social network characteristics on factuality for the 9 datasets for checking robustness of results.}
\end{figure}

\newpage
\subsection*{Interactions}

Through plotting the interactions of follower count with all other metrics, certain differences in the relationships between these metrics and factuality levels across users with varying follower counts have been observed. Specifically, our analysis reveals that in case of users with fewer followers the effect of the followed account count diminishes on high factuality, whereas those with a larger follower base show an increased likelihood of high factuality with more followed accounts (Figure 11, panel a). Thus, the effect of followed account count on high factuality is nonexistent among those with smaller follower base, while positive among those with larger follower base. For users in the low factuality group, we observe a similar trend across both follower count categories: as users follow more accounts, they become less likely to belong to the low factuality group. This effect is slightly stronger for users with a higher follower count. Among those with fewer followed accounts, popular users are more likely to have low factuality, while in case of those who follow many accounts, there is no significant difference in their likelihood of low factuality by popularity.

In relation to average daily tweet count our analysis (Figure 11, panel b) shows that as users increase their tweet activity, they are less likely to belong to the high factuality category and more likely to belong to the low factuality category, regardless of their follower count. Nevertheless, in the low factuality group, this relationship is less pronounced among users with fewer followers.

Regarding the relationship between factuality and days since registration (Figure 11, panel c) we show that as users spend more time registered on the platform, they are more likely to belong to the high factuality category and less likely to belong to the low factuality category, and this is consistent in direction and magnitude across users, regardless of their follower count. However, the positive effect on high factuality is slightly stronger among users with higher follower counts. For the low factuality group, we observe a negative relationship with days since registration, with a marginally stronger effect among those with more followers. It shows that among younger users popular ones are more likely to have lower factuality compared to less popular ones. However, among older users there is no difference in the likelihood of factuality between more and less popular users.

In examining average daily tweet count across different levels of followed account count and its association with factuality levels (Figure 11, panel d) we find that among those with few accounts followed the average tweet count does not affect the likelihood of belonging to the high factuality group. However, an increase in tweet activity is associated with an increased likelihood of belonging to the high factuality group among those who follow more accounts. For the low factuality group, tweet count is negatively associated with low factuality and this effect is stronger among those who follow more accounts. There is no significant difference in the likelihood of belonging to the low factuality group among those who tweet infrequently, however, among those who tweet frequently, users following few accounts are way more likely to have low factuality than users who follow more accounts. Figure 11, panel d corresponds to Figure 6, panel a in the main text.

Examining the interaction between days since registration and tweet count in relation to factuality levels reveals consistent trends across users with different average daily tweet counts (Figure 11, panel e). Our analysis demonstrates that as the registration period lengthens, users are more likely to exhibit higher factuality levels, and less likely to exhibit lower factuality levels, irrespective of their tweet activity.

Our findings regarding the interaction between days since registration and followed account count (Figure 11, panel f) indicate that among users who follow fewer accounts, those who have been registered on the platform for a longer duration are more likely to be in the high factuality category, while among those who follow more accounts, the trend is the opposite: the longer they are on the platform, the less likely that they belong to the high factuality group. The relationship between days since registration and low factuality does not vary based on followed account count. The relationship is consistently negative between days since registration and low factuality, indicating that longer registration periods are associated with a decreased likelihood of low factuality, however, this effect is somewhat more pronounced among users who follow fewer accounts. Figure 11, panel f corresponds to Figure 6, panel b in the main text.

\begin{figure}[h]
\centering
\includegraphics[width=1\textwidth]{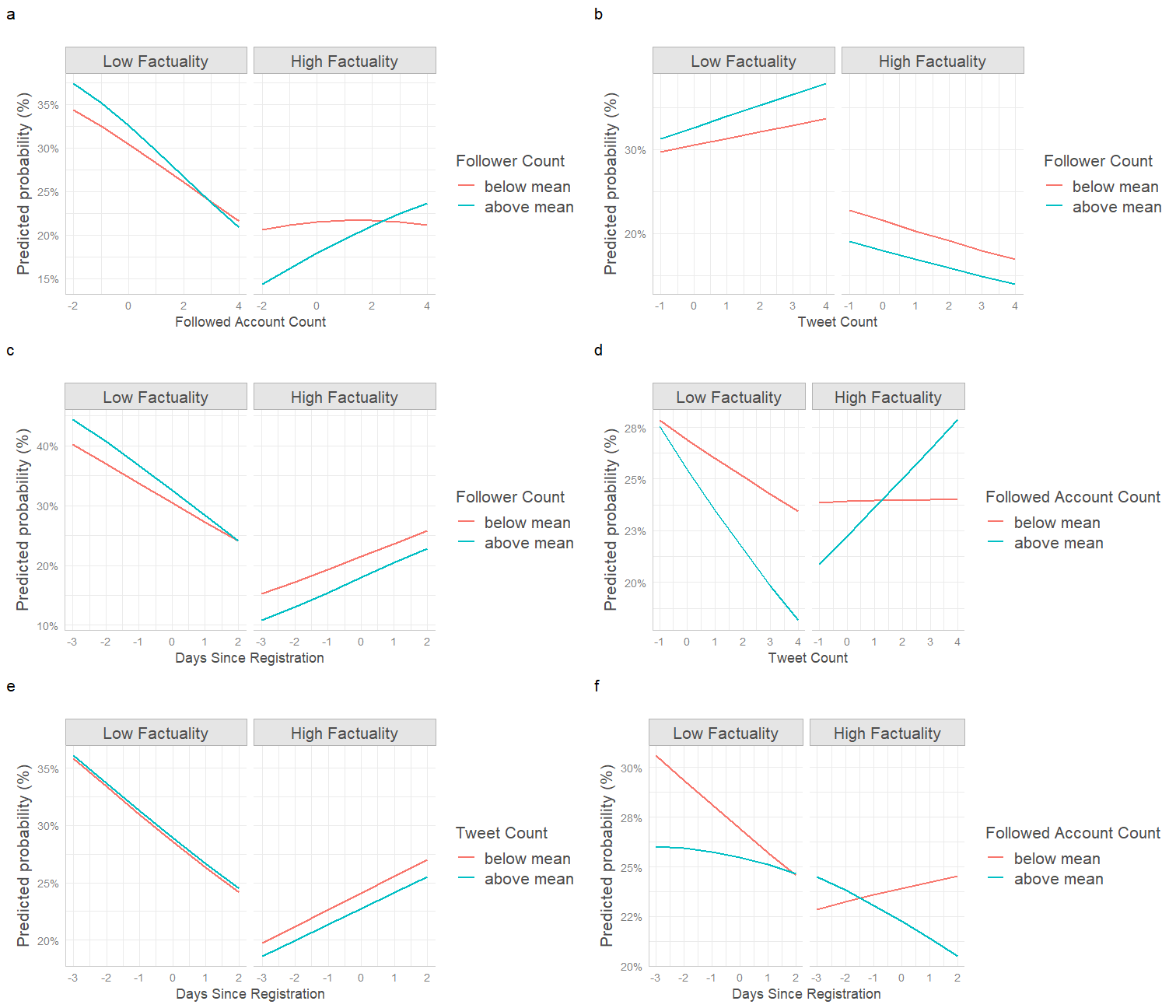}
\caption{Interactions between various social network metrics in relation to belonging to low and high factuality groups.}
\end{figure}

\clearpage
\newpage
\subsection*{Pseudo-R (McFaddens R)}

Pseudo-R (McFaddens R) obtained from shuffled datasets against those derived from the real data to demonstrate that the Pseudo-R obtained from our regression exceeds that expected by random chance.

\begin{figure}[h]
\centering
\includegraphics[width=1\textwidth]{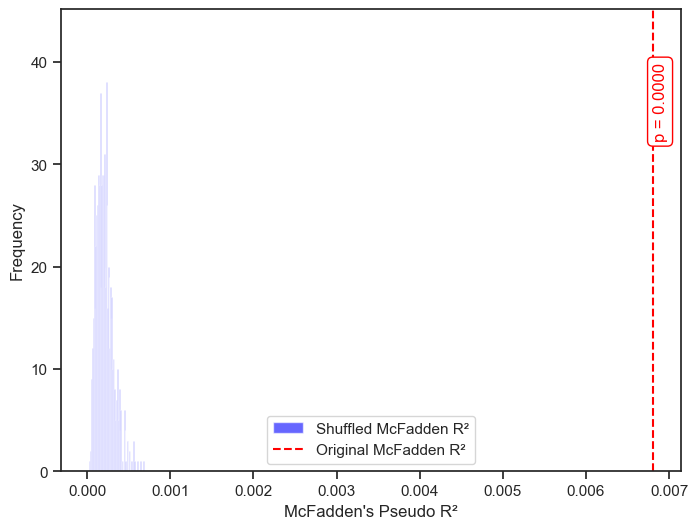}
\caption{Pseudo-R (McFaddens R) using the four social network metrics on the real dataset (red dotted line) compared to the accuracy scores obtained on the 1000 randomised datasets (blue bars).}
\end{figure}

\end{document}